\documentclass{nature}

\usepackage{amsmath,amssymb}
\usepackage{xcolor}
\usepackage{times}
\usepackage{graphicx}
\usepackage[utf8]{inputenc}

\usepackage{tikz}

\usepackage[utf8]{inputenc}

\title{Time-(ir)reversibility in active matter: from micro to macro}

\author{J. O'Byrne$^1$, Y. Kafri$^{2}$, J. Tailleur$^1$, F. van Wijland$^1$}

\date{\today}
\usepackage{hyperref}
\newcommand\bfr{{\bf r}}
\newcommand\bfv{{\bf v}}
\newcommand\bfu{{\bf u}}
\newcommand\bfp{{\bf f}_p}
\newcommand\bfeta{{\boldsymbol \eta}}
\newcommand\bfB{{\bf B}}
\newcommand\bfJ{{\bf J}}
\newcommand\bff{{\bf f}}
\newcommand\bfF{{\bf F}}
\usepackage{siunitx}
\usepackage{enumitem}
\definecolor{mygray}{rgb}{0.9,0.9,.9}
\usepackage[many]{tcolorbox}
\tcbset{width=\textwidth,boxrule=0pt,colback=mygray,arc=2pt,auto outer arc,left=0pt,right=0pt,boxsep=5pt}
\newcommand\pluspart {\tikz[baseline=-.5ex,scale=0.9]{\filldraw[blue!40!white](2.5,0) circle (1ex) node {\color{black} $\boldsymbol{+}$};}}

\newcommand\minuspart {\tikz[baseline=-.5ex,scale=0.9]{\filldraw[red!40!white](2.5,0) circle (1ex) node {\color{black} $\boldsymbol{-}$};}}

\newcommand\alt{1}
\newcommand\entroprop{1}
\newcommand\selfconsistentpot{1}
\begin{document} 

\maketitle 
\begin{affiliations}

\item Universit\'e de Paris, MSC, UMR 7057 CNRS, 75205 Paris, France

\item Department of Physics, Technion, Haifa, 32000, Israel

\end{affiliations}

\begin{abstract}
  Active matter encompasses systems whose individual consituents
  dissipate energy to exert propelling forces on their
  environment. This rapidly developing field harbors a dynamical
  phenomenology with no counterpart in passive systems. The extent to
  which this is rooted in the breaking of time-reversibility has
  recently triggered an important theoretical and experimental
  activity which is the focus of this review. Building on recent
  progress in the field, we disentangle the respective roles of the
  arrow of time and of the non-Boltzmann nature of steady-state
  fluctuations in single- and many-body active systems.  We show that
  effective time-reversible descriptions of active systems may be
  found at all scales, and discuss how interactions, either between
  constituents or with external operators, may reveal the
  non-equilibrium nature of the microscopic source of energy. At a
  time when the engineering of active materials appears within our
  reach, this allows us to discuss to which extent methods stemming
  from equilibrium statistical mechanics may guide us in their design.
\end{abstract}

\section{Introduction}

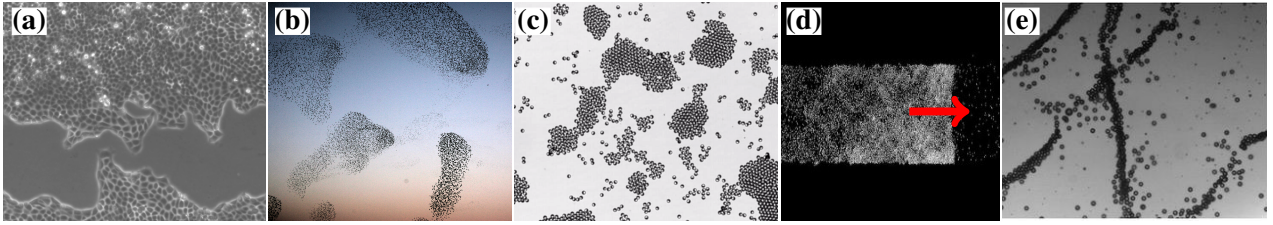
\begin{figure}
    \centering
    \begin{tikzpicture}
    \def\x{3.5}
\path (0.1,0) node {    \includegraphics[totalheight=2.9cm]{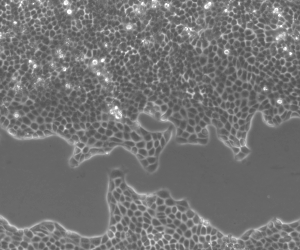}};
\path (3.5,0) node {    \includegraphics[totalheight=2.9cm]{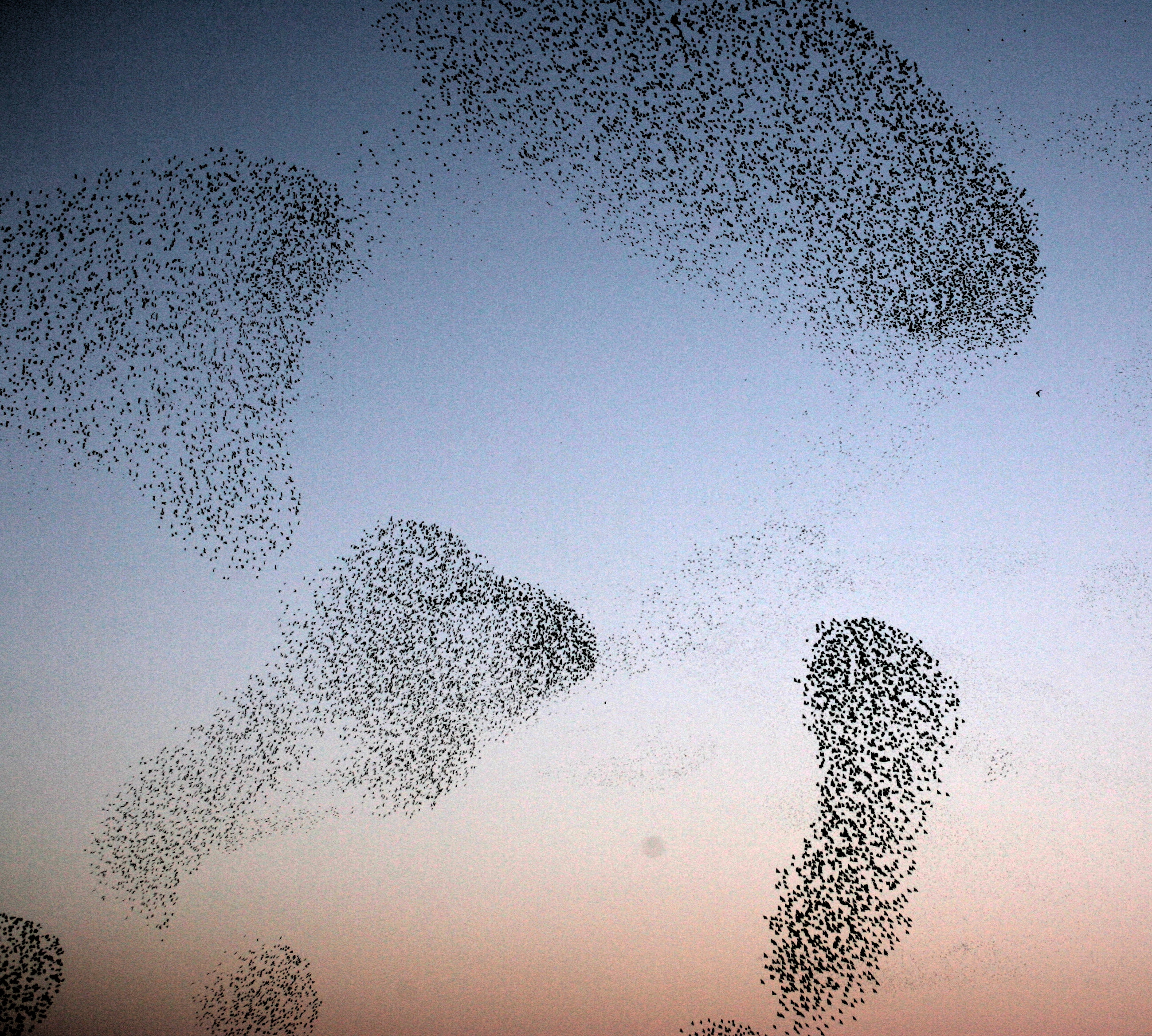}};
\path (6.9,0) node {    \includegraphics[totalheight=2.9cm]{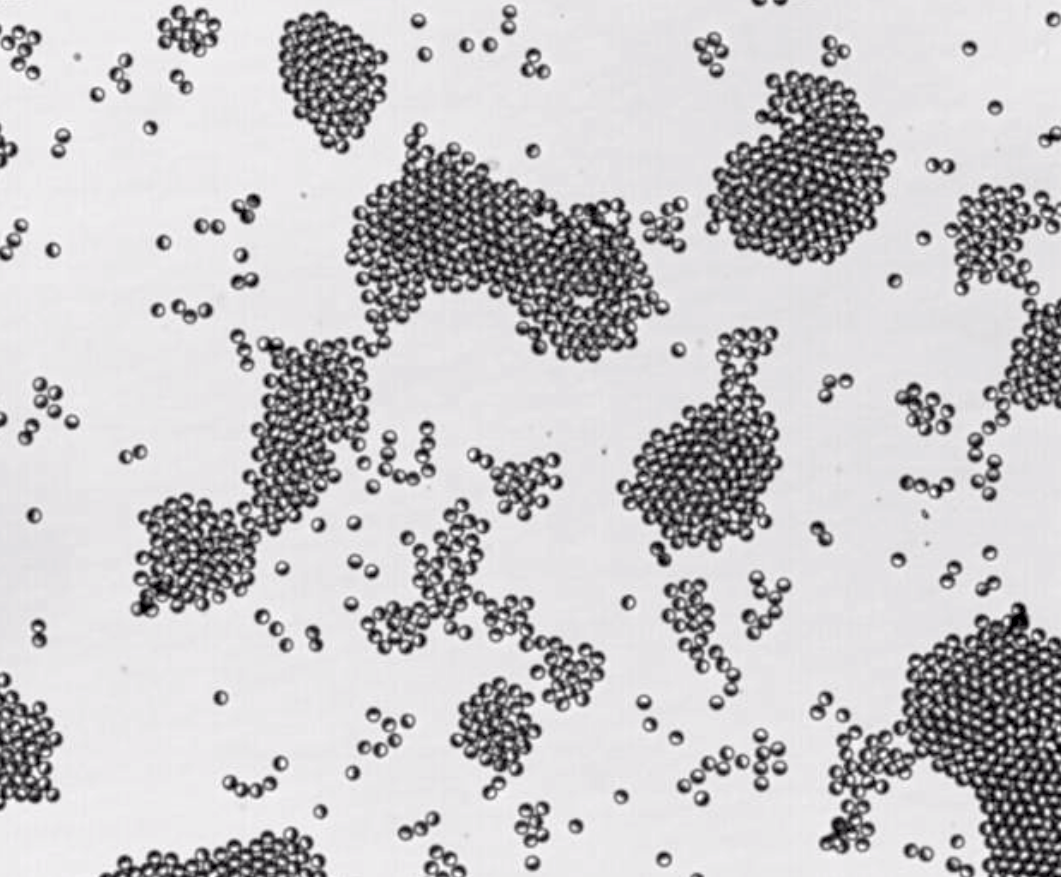}};
\path (10.15,0) node {    \includegraphics[totalheight=2.9cm]{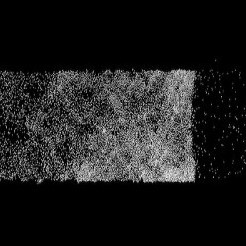}};
\draw[red,->,line width = 3pt] (9.7+.7,0) -- +(.8,0);
\path (13.4,0) node {    \includegraphics[totalheight=2.9cm]{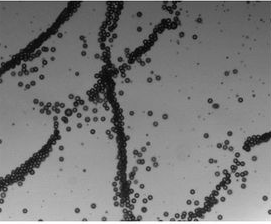}};
\begin{scope}[xshift=0cm,yshift=.2cm]
\def\x{.3}
\filldraw[white] (-1.225-\x,.825) rectangle (-.775-\x,1.175);
\draw (-1-\x,1) node {\small{\bf (a)}};
\end{scope}
\begin{scope}[xshift=3.5cm,yshift=.2cm]
\def\x{.3}
\filldraw[white] (-1.225-\x,.825) rectangle (-.775-\x,1.175);
\draw (-1-\x,1) node {\small{\bf (b)}};
\end{scope}

\begin{scope}[xshift=6.7cm,yshift=.2cm]
\def\x{.3}
\filldraw[white] (-1.225-\x,.825) rectangle (-.775-\x,1.175);
\draw (-1-\x,1) node {\small{\bf (c)}};
\end{scope}

\begin{scope}[xshift=10.3cm,yshift=.2cm]
\def\x{.3}
\filldraw[white] (-1.225-\x,.825) rectangle (-.775-\x,1.175);
\draw (-1-\x,1) node {\small{\bf (d)}};
\end{scope}

\begin{scope}[xshift=13.2cm,yshift=.2cm]
\def\x{.3}
\filldraw[white] (-1.225-\x,.825) rectangle (-.775-\x,1.175);
\draw (-1-\x,1) node {\small{\bf (e)}};
\end{scope}
    \end{tikzpicture}
    \caption{Active matter systems are commonly found in nature (a-b) and engineered in the lab (c-e). 
    {\bf (a)} Monolayer of cells invading an artifical wound~\cite{poujade2007collective} (Copyright (2007) National Academy of Sciences). {\bf (b)} Bird flocks undergoing collective motion~\cite{bialek2012statistical} (Credits COBBS Lab, Institute for Complex Systems, Rome). {\bf (c)} Interrupted motility-induced phase separation in self-propelled colloids~\cite{van2019interrupted} (Courtesy of O. Dauchot). {\bf (d)} Non-linear travelling wave of Quincke colloidal rollers~\cite{Bricard:2013:Nature} (Courtesy of D. Bartolo and A. Morin). {\bf (e)} Dynamic lane formation and breaking by self-propelled droplets~\cite{thutupalli2018flow}. }
    \label{fig:collective}
\end{figure}

Active matter describes systems whose fundamental constituents
dissipate energy to exert self-propelling forces on the
environment. From birds to bacteria, from cells to molecular motors,
the biological world is filled with active entities. Indeed, active
matter was initially strongly driven by its biophysical applications
(Fig~\ref{fig:collective}a-b). Since then, a wealth of synthetic
active systems have been engineered in the lab, which now pave the way
towards the engineering of synthetic active materials
(Fig~\ref{fig:collective}c-e). However, many fundamental questions
need to be addressed before one may reach the level of complexity
relevant for material design or to decipher the physical laws involved
in biological processes, from morphogenesis to the large-scale
organization of complex ecosystems.

In addition to its potential implications for biophysics and material
design, the upsurge of interest in active matter also stems from its
rich dynamical phenomenology, which is largely without counterpart in
passive systems and has been extensively reviewed
previously~\cite{cates2012diffusive,Vicsek:2012:PhysRep,romanczuk2012active,Marchetti:2013:RMP,gonnella2015motility,Cates:2015:ARCMP,Bechinger:2016:RMP}.
This emerging physics stems from a microscopic nonequilibrium drive at
the particle level that leads to system-dependent violations of two
hallmark features of thermal equilibrium: time-reversal symmetry (TRS)
and Boltzmann statistics.
For instance, the lack of TRS, when accompanied by the spontaneous
emergence of macroscopic currents, is at the root of flocking
behaviours, whether observed in groups of animals
(Fig.~\ref{fig:collective}b) or engineered in synthetic self-propelled
colloids (Fig.~\ref{fig:collective}d). Nothing prevents, however, the
orientations of the birds within a flock to display an
equilibrium-like statistics~\cite{Mora2016NatPhys}. Conversely, the
emergence of self-assembled clusters in the absence of attractive
forces shown in Fig.~\ref{fig:collective}c transcends the constraints
imposed by the Boltzmann weight. No manifest arrow of time is,
however, observed in the corresponding motility-induced phase
separation.\if{se experiments.}\fi
The aim of this review is to disentangle the respective roles of the
violations of TRS and of Boltzmann statistics in active matter and to
discuss the vast literature which touches on these questions.
At a time when the design of active materials has become a practical
challenge, the existence of an effective TRS---and of its accompanying
equilibrium toolbox--- has become a central question. 

This review is intended both for newcomers in the field of active
matter as well as for condensed-matter and biophysics specialists. As
such, it relies on general concepts of statistical mechanics and
stochastic processes. We start by discussing the Langevin description
of active particles in section~\ref{sec:non-interacting}, which we
contrast with the Brownian dynamics of colloidal particles. We explain
the meaning of TRS in this stochastic context and how TRS violations
can be measured using the entropy production rate introduced in
stochastic thermodynamics. We discuss the ambiguities that naturally
emerge when characterizing the TRS of coarse-grained systems, which we
loosely understand as integrating out some of the degrees of
freedom. We illustrate this discussion by considering a free active
particle whose position and orientation are recorded over time. TRS
violation is demonstrated by a non-vanishing entropy production rate
equal to the energy dissipated by the active force powering the
motion. Measuring solely the particle position, however, restores TRS
and leads to a vanishing entropy production rate.

We then turn in section~\ref{sec:obstacles} to the case of
non-interacting active particles subjected to external potentials. We
discuss how non-Boltzmann features emerge in this context. In
particular, we compare the non-local dependency of the steady-state
distribution on the external potential for different types of active
particles. In addition, external potentials also lead to TRS
violations, under the form of non-vanishing entropy production rates
and steady-state currents. Conversely, we discuss the conditions under
which an effective thermal equilibrium regime is recovered or a TRS is
restored.

Finally, section~\ref{sec:interactions} discusses the case of
interacting active particles. We focus in particular on
Motility-Induced Phase Separation (MIPS) whose macroscopic dynamics
resembles an equilibrium phase separation despite a microscopic
nonequilibrium drive at the particle level, which makes it interesting
from a TRS perspective. We show how TRS violations can nevertheless be
identified using field-theoretical methods and how they impact the
construction of the phase diagram. We also discuss cases in which TRS
is exactly restored at the macroscopic scale, either by considering
specific classes of active systems endowed with tactic interactions or
in the small-persistence-time limit.

\section{TRS of a single active-particle: in and out of equilibrium}
\label{sec:non-interacting}

\subsection{Sources and sinks of energy in active matter.}
Let us consider the simplest model of an active particle
\begin{equation}\label{eq:dynact}
   m\ddot \bfr = - \gamma \dot \bfr + {\bf f}_p(t) -\nabla V(\bfr) + \sqrt{2 \gamma^2 D} \bfeta(t)
\end{equation}
Here, $\bfr$ is the position of the particle, $-\gamma \dot\bfr $ a
viscous damping term, ${\bf f}_p$ the self-propulsion force, $V(\bfr)$
an external potential, and $\bfeta$ a Gaussian white noise that may
represent thermal noise or other fluctuation sources. The
self-propulsion force ${\bf f}_p$ is a defining feature of active
matter: it is a time-dependent, non-conservative force, endowed with a
particle-dependent stochastic dynamics, and powered by an irreversible
consumption of energy. The presence of self-propulsion forces both
drive active systems out of thermal equilibrium and distinguish active
matter from other nonequilibrium systems, such as glasses or boundary
driven fluids whose bulk dynamics is of equilibrium nature. The energy
source powering ${\bf f}_p$ varies from system to system: it can be a
local field, as in Janus self-phoretic
colloids~\cite{howse2007self,Palacci2010PRL,palacci2013living} or
bacteria~\cite{berg2008coli,cates2012diffusive}, or it can stem from a
global external field, as for vibrated
grains~\cite{Deseigne2010PRL}, Quincke
rollers~\cite{Bricard:2013:Nature}, or bi-electric colloids under AC
fields~\cite{nishiguchi2015mesoscopic,yan2016reconfiguring,van2019interrupted}. The
dynamics of ${\bf f}_p$ can be equally diverse and can, for instance,
include rotational diffusion or tumbles of its orientation, as well as
fluctuations in its amplitude.

\begin{figure}
\centering
    \includegraphics{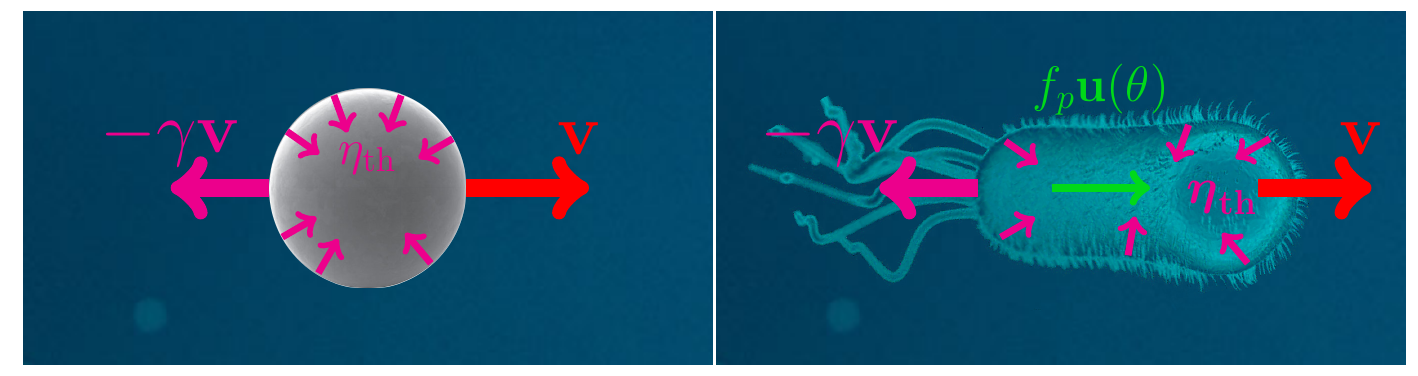}
    \caption{Colloidal dynamics resulting from interactions with fluid
      molecules ({\bf left}). The mean resulting force is a drag
      $-\gamma \bfv$, which dissipates energy by opposing motion,
      whereas fluctuations around it, described by the Gaussian white
      noise $\eta_{\rm th}$, inject energy into the colloid's
      dynamics. When the fluid is in equilibrium, injection and
      dissipation are related by the fluctuation-dissipation
      theorem. On the contrary, the propulsion of a bacterium ({\bf
        right}), stems from the consumption of a separate energy
      source, which is disconnected from the dissipation. In the
      steady state, the energy balance between sources and sinks of
      energy is given by
      $\langle \bfp \cdot \dot \bfr \rangle + \frac{ d \gamma^2 D}{m}
      = \gamma \langle \dot \bfr ^2 \rangle$, which can be derived
      from Eq.~\eqref{eq:dynact} in $d$ space dimensions by computing
      the time derivative of $m\dot \bfr^2/2$.  The dissipation
      $w_p=\langle \bfp \cdot \dot \bfr \rangle$ measures the energy
      per unit time injected by the self-propulsion force into the
      system; it violates the equipartition
      $\frac 1 2 m \langle\dot \bfr^2\rangle = \frac 1 2 d \gamma D$ that
      would otherwise be satisfied for an equilibrated colloid. It is
      the microscopic signature of activity.  }\label{fig:bactos}
\end{figure}

A natural question is whether one could subsume $\bfp(t)$ and
$\sqrt{2 \gamma^2 D} \bfeta(t)$ into a single noise source
$\tilde \bfeta(t)$ and treat Eq.~\eqref{eq:dynact} as an effective
equilibrium Langevin dynamics. The answer comes from thermodynamics:
the main difference between passive and active particles lies in the
injection of energy. For passive particles, the fluid is responsible
both for the dissipation and the injection of energy
(Fig.~\ref{fig:bactos}, left). The former results from the average
force exerted by the fluid on the particle, $-\gamma \dot \bfr$, with
a corresponding dissipated power $-\gamma {\dot\bfr}^2$. On the
contrary, the latter stems from fluctuations of the fluid forces
around their mean, $\sqrt{2 \gamma^2 D} \bfeta(t)$, which inject a
mean power $d \gamma^2 D/m$. When the fluid is at equilibrium,
injection and dissipation of energy are related by the celebrated
fluctuation-dissipation theorem through $D=k T/\gamma$. For an active
particle, on the contrary, most of the injection of energy comes from
the self-propulsion force and it is thus physically disconnected from
dissipation: $\tilde \bfeta(t)$ would not satisfy a
fluctuation-dissipation relation (See Fig.~\ref{fig:bactos},
right). Note that this discussion assumes that ${\bf f}_p$ is not a
Gaussian white noise, otherwise it could indeed be absorbed in
$\bfeta(t)$ at the cost of a simple quantitative shift of the
temperature. This observation is the motivation for most models of
active particles considered in the literature that rely on
variations of Eq.~\eqref{eq:dynact} with noises and drags not related
by (generalized) Stokes-Einstein relations.

Since most active systems live at low Reynolds numbers, the overdamped
limit of the Langevin dynamics is often assumed, although the
role of inertia has recently attracted
interest~\cite{attanasi2014information,manacorda2017lattice,de2019flow,mandal2019motility,dai2020phase,lowen2020inertial}
. Three standard models of active particles are described by the
dynamical equation
\begin{equation}\label{eq:dynod}
    \dot \bfr = \bfv_p -\mu \nabla V(\bfr) +\sqrt{2 D} \bfeta(t)
\end{equation}
where $\mu=\gamma^{-1}$ is the particle mobility and
$\bfv_p=\mu {\bf f}_p$ its self-propulsion velocity. For Active
Brownian particles (ABPs) and Run-and-Tumble particles (RTPs), the
magnitude of $\bfv_p$ is constant, while its orientation undergoes
either rotational diffusion or random, Poisson-distributed
reorientation events. On the contrary, an Active Ornstein-Uhlenbeck
particle (AOUP) has a self-propulsion velocity whose modulus
fluctuates and is given by an Ornstein-Uhlenbeck process
$\dot \bfv_p = - \bfv_p/\tau + \sqrt{2 D \tau^2} \boldsymbol\xi$, with
$\boldsymbol\xi$ a Gaussian white noise with unit amplitude.  Using
angular brackets to denote averages over the dynamics of the
self-propulsion velocities, the temporal correlation of
$\langle \bfv_p(t) \bfv_p(0) \rangle$ can be equal in the three
models. The physics of their self-propulsion, however, is quite
different: the probability distribution of $\bfv_p$ is Gaussian and
peaked around ${\bf 0}$ for AOUPs, whereas it is uniform on a sphere
of fixed radius for ABPs and RTPs.

Note that the discussion above for the translational degree of freedom
$\bfr$ can be extended to rotational ones. In particular, this leads
to a class of particles called spinors, in which self-propulsion speed
is replaced with a self torque or, more generally, the rotational
noise and damping do not satisfy a Stokes-Einstein relation. These
systems have been studied both at the
theoretical~\cite{nguyen2014emergent,yeo2015collective,
  van2016spatiotemporal, goto2015purely, liebchen2016pattern,
  aragones2016elasticity} and experimental~\cite{kokot2017active,
  sabrina2018shape, soni2019odd, brooks2019shape} levels.

An immediate consequence of the absence of the Stokes-Einstein
relation is that, in the presence of an external potential $V$, the
stochastic dynamics of an active particle does not necessarily lead to
a steady state given by a Boltzmann weight
$P_{\rm eq}({\bfr})\propto\exp[-\beta V(\bfr{})]$. This leads to a
host of interesting phenomena that we partly review in
Section~\ref{sec:obstacles}. Note, however, that while the Boltzmann
weight may be the textbook definition of equilibrium statistical
mechanics, a broader definition of equilibrium is based on dynamics;
it equates with TRS in the steady state, under the form of a
detailed-balance relation. We now discuss the origin of the latter and
its rather subtle fate for active particles.

\subsection{Equilibrium \& time-reversal symmetry.}
The dynamics of a system is said to be time-reversal symmetric if
observing a trajectory or its time-reversal are equally likely:
$P[\{\bfr(\tau), 0 \leq \tau\leq t\}]=P[\{\bfr(t-\tau), 0 \leq
\tau\leq t\}]$, which, using the definition of conditional
probability, can equally be written as
\begin{equation}\label{eq:TRScolloid}
   P[\{\bfr(\tau), 0 < \tau\leq t\}|\bfr(0)] P_{\rm s}(\bfr(0))  =  P[\{\bfr(t-\tau), 0 < \tau\leq t\}|\bfr(t)] P_{\rm s}(\bfr(t))\;.
\end{equation}
Here $P_{\rm s}(\bfr)$ is the (stationary) probability to sample a
configuration $\bfr$ whereas
$P[\{\bfr(\tau), 0 \leq \tau\leq t\}|\bfr(0)]$ is the probability of
any trajectory $\{\bfr(\tau)\}$ starting from $\bfr(0)$. In the
following, we restrict ourselves to Markovian dynamics for simplicity,
even though the discussion can be generalized to non-Markovian
processes. $P[\{\bfr(\tau), 0 < \tau\leq t\}|\bfr(0)]$ is then the
average probability that a trajectory passing through $\bfr(0)$ at 0
will follow $\{\bfr(\tau)\}$. Microscopically,
Eq.~\eqref{eq:TRScolloid} is granted by the laws of classical
mechanics and directly applies if $\bfr$ describes both a colloidal
particle and the surrounding fluid molecules. Mesoscopically, however,
when $\bfr$ solely describes the colloidal particle,
Eq.~\eqref{eq:TRScolloid} requires a statistical hypothesis on the
distribution of the fluid degrees of freedom which have been
coarse-grained out (See Box \textit{Microscopic vs statistical
  reversibility}). In equilibrium, this is given by Boltzmann
microcanonical hypothesis. For more general systems, coarse-graining
out the fluid degrees of freedom need not lead to dynamics obeying a
statistical time-reversal symmetry. A whole set of theoretical tools
have thus been developed to test whether stochastic dynamics such as
Eq.~\eqref{eq:dynact} or~\eqref{eq:dynod} satisfy TRS or not (See
Box~\textit{Identifying time-reversal symmetry breaking}).

\begin{tcolorbox}
{\bf Microscopic vs statistical reversibility.}  Consider a colloidal
particle embedded in an equilibrated fluid. When considering the full
system, \{Fluid + Colloid\}, time-reversibility is given by
Eq.~\eqref{eq:TRScolloid}, where $\bfr(s)$ describes the joint degrees
of freedom of the colloid and of the fluid molecules. The
time-reversibility of Hamilton's equations of motion ensures that $
P[\{\bfr(\tau), 0 \leq \tau\leq t\}|\bfr(0)]=P[\{\bfr(t-\tau), 0 \leq
  \tau\leq t\}|\bfr(t)]$. This directly leads to $P_{\rm
  s}(\bfr(0))=P_{\rm s}(\bfr(t))$: the probability is conserved along
the trajectory, as guaranteed by Liouville theorem.

If $\bfr(s)$ now refers to the colloid only, then
$P[\{\bfr(\tau), 0 \leq \tau\leq t\}|\bfr(0)]$ is given by the measure
of all initial conditions of the fluid molecules which are compatible
with the trajectory of the colloid. Assuming microcanonical
equilibrium, a scale separation between the fluid molecules and the
colloid, and denoting by $E(\bfr)$ and $E_{\rm tot}$ the energies of
the colloid and of the total system, the measure of these initial
conditions is given by
$\Omega_{\rm traj}/\Omega_{\rm fl}[E_{\rm tot}-E(\bfr(0))]$. Here,
$\Omega_{\rm fl}[E_{\rm tot}-E(\bfr(0))]$ is the phase-space volume of
all fluid configurations at energy $E_{\rm tot}-E(\bfr(0))$ and
$\Omega_{\rm traj}$ is that of the initial conditions compatible with
the colloid trajectory. Conversely, the probability of the
time-reversed trajectory is given by
$\Omega_{\rm traj}/\Omega_{\rm fl}[E_{\rm tot}-E(\bfr(t))]$, which now
differs from that of the forward trajectories if the energy of the
colloid has changed. The time-reversal condition~\eqref{eq:TRScolloid}
then simplifies into
\begin{equation}\label{eq:BW}
     P_{\rm s}(\bfr(0)) \exp[E(\bfr(0))/kT] = P_{\rm s}(\bfr(t)) \exp[E(\bfr(t))/kT]
\end{equation}
where we have used that $\Omega_{\rm fl}(E_{\rm tot}-E(\bfr))\propto
\exp(-E(\bfr)/kT)$ with $T$ the temperature of the fluid, defined as
$T^{-1}=k\left.\frac{\partial \log\Omega_{\rm fl}}{\partial
  E}\right|_{E_{\rm tot}}$. Equation~\eqref{eq:BW} enforces the Boltzmann weight as a
steady-state distribution for the colloidal particle.

Unlike the microscopic reversibility of Hamilton's equations of
motion, the statistical reversibility of the colloid's dynamics
requires a statistical hypothesis for the fluid molecules, which is
here given by the microcanonical postulate. Another distribution for
the fluid molecules would typically have led to a violation of
Eq.~\eqref{eq:TRScolloid}: statistical irreversibility for the
colloidal dynamics thus generically emerges from a microscopically
reversible dynamics once the bath's degrees of freedom have been
coarse-grained out. 
\end{tcolorbox}

\begin{tcolorbox}[breakable]
{\bf Identifying time-reversal symmetry breaking.}
The question of TRS and its breakdown arises in many different settings and formalisms, whether using Langevin dynamics or jump processes, for underdamped or overdamped dynamics, at the level of particles or fields. Its identification relies on an equally diverse set of tools~\cite{gardiner1985handbook,van1992stochastic}, which we illustrate for the simple setting of an overdamped Brownian dynamics in the presence of a force field $\bf{F}(\bfr)$:
\begin{equation}
\dot \bfr = \mu\bf{F}(\bfr) + \sqrt{2D}\bfeta \;,
\label{eq:underLang}
\end{equation}
where $\bfeta$ is a unit Gaussian white noise and $D$ and $\mu$ positive constants. For such Markovian dynamics, the TRS defined in Eq.~\eqref{eq:TRScolloid} can be read at several levels.
\begin{itemize}[leftmargin=.4cm]
    \item TRS amounts to an equality in the steady state of the joint two-time probability density $p(\bfr_1,t ; \bfr_2,0)=p(\bfr_2,t ; \bfr_1,0)$. This generalizes to any pairs of observables $A(\bfr)$ and $B(\bfr)$ where TRS implies $C_{AB}(t)=C_{BA}(t)$ with $C_{AB}(t)=\langle A(t) B(0)\rangle$. In experiments or simulations, measuring a non-vanishing $\Delta_{AB}(t)\equiv [C_{AB}(t)-C_{BA}(t)]/2$ offers a low-dimensional \textit{sufficient} condition to prove a violation of TRS. $\Delta_{AB}(t)$ is the time anti-symmetric part of the correlation function $C_{AB}(t)$, an idea which can be generalized to other observables~\cite{maes2020frenesy}.
    
    \item The probability density $P(\bfr,t)$ evolves according to a Fokker-Planck equation $\partial_t P=-H P$. TRS is equivalent to a symmetry of the Fokker-Planck operator: $H^{\dagger}=P_S^{-1}H P_S$, where $P_S$ is the stationary measure and $H^{\dagger}$ is the adjoint of $H$ (for the $L^2$ scalar product). In particular, this implies the existence of a basis in which $H$ becomes Hermitian and that it has a real spectrum~\cite{risken1996fokker}. For overdamped Brownian dynamics with additive noise, this has been used to endow $H$ with a supersymmetric structure akin to that of Schr\"odinger's equation~\cite{tanase2003statistical,witten1982supersymmetry}. 

    \item The Fokker-Planck equation can be read as a conservation
      equation for a probability current
      ${\bf J}(\bfr)=\mu \bfF(r) P(\bfr)-D \nabla P(\bfr)$. TRS is
      equivalent to the vanishing of $\bfJ{}$ in the steady state. The
      experimental measurement of such a high dimensional object is
      difficult, but its projection on a finite set of relevant
      degrees of freedom has recently been used to sample TRS
      breakdown in biological systems~\cite{gladrow2016broken,Battle2016Science,gnesotto2018broken}.

\item Another measure of TRS breakdown is obtained by computing:
\begin{equation*}
    \Sigma(0,t_f)=\int D[\{\bfr(t)\}] P[\{\bfr(t)\}] \hat \Sigma [\{\bfr(t)\}] =  \int D[\{\bfr(t)\}] P[\{\bfr(t)\}] \log \frac{P[\{\bfr(t)\}]}{{P[\{\bfr(t_f-t)\}]}}\,.
\end{equation*}
Mathematically, $\Sigma$ is the Kullback-Leibler divergence between
the probabilities of backward and forward paths and $\hat \Sigma$ can
be seen as a `path-wise entropy production'~\cite{seifert2005entropy}.
For dynamics~\eqref{eq:underLang}, $\Sigma$ can be decomposed between
the heat transferred to the thermal bath and the variation of the
Gibbs-Shannon entropy of the measure
$P(\bfr,t)$\cite{hatano2001steady,seifert2005entropy}. The positivity
of $\Sigma$ generalizes the second law of Thermodynamics to
microscopic systems described by Langevin equations. In turn, the
entropy production rate, defined as:
\begin{equation}\label{eq:defsigma}
  \sigma\equiv \lim_{t_f\to\infty} \frac{1}{t_f} \Sigma(0,t_f)\;,
\end{equation}
is a direct measurement of the irreversibility of the
dynamics~\cite{maes1999fluctuation,lebowitz1999gallavotti,kurchan1998fluctuation}
that is commonly used in non-equilibrium statistical mechanics, even
in the absence of any connection to
Thermodynamics~\cite{dabelow2019irreversibility,maes2020local}.
\end{itemize}
Of course, the four criteria discussed above are not independent from
each other. For instance, for dynamics~\eqref{eq:underLang}, the
steady-state entropy production rate is connected to the probability
current $\bfJ(\bfr)$ defined above through:
\begin{equation*}
\sigma =\frac{\mu}{D} \int d \bfr \frac{ \bfJ^2(\bfr) }{P_s(\bf r)}\;.
\end{equation*}
From the knowledge of the dynamics, conditions can be derived for TRS
to be violated. For dynamics~\eqref{eq:underLang}, irreversibility
requires the existence of a closed loop ${\cal C}$ such that:
\begin{equation*}
    \oint_{\cal C} \bfF(\bfr) \cdot d\bfr \neq 0     
\end{equation*}
This criterion generalizes the Kolmogorov criterion derived for Markov
chains~\cite{van1992stochastic}. Mathematically, ${\bfF}$ needs to
contain a curl or a harmonic
part~\cite{jiang2004mathematical}. Conversely, TRS imposes a direct
relation between the force field $\bf F$ and the steady-state
probability: $\mu \bfF = D \nabla \log P_s$. All the discussion above
generalizes---under more complex forms---to other dynamics, for
instance involving variables which are odds under time-reversal
symmetry~\cite{van1992stochastic,gardiner1985handbook}.
\end{tcolorbox}

\subsection{TRS in active matter.}
\label{subsec:TRSnointeraction}
For equilibrium systems, TRS is ensured irrespectively of the degree
of coarse graining and holds both for microscopic and coarse-grained
mesoscopic descriptions. For active systems, on the contrary, this
question is more subtle and the possible existence of TRS depends on
the degrees of freedom which are being considered. This makes its
characterization---in particular using an entropy production rate
defined in the spirit of Eq.~\eqref{eq:defsigma}---a somewhat ambiguous
task, which has attracted a lot of interest both
theoretically~\cite{Ganguly2013PRE,guo2014probing,Fodor:2016:PRL,Nardini:PhysRevX.7.021007,mandal2017entropy,caprini2018comment,roldan2018arrow,shankar2018hidden,dadhichi2018origins,szamel2019stochastic,dabelow2019irreversibility,borthne2020time,martin2020aoup}
and
experimentally~\cite{Mizuno2007Science,wilhelm2008out,robert2010vivo,fodor2015activity,fodor2016nonequilibrium,gladrow2016broken,Battle2016Science,gnesotto2018broken}.

To illustrate this, consider a zero-Reynolds swimmer at position
$\bfr(t)$ moving through a fluid thanks to the displacement of some
degrees of freedom ${\bf x}_i(t)$. (Think about the motion of a
flagellum.) The solution of Stokes equation will lead to a flow
$\bfu(t)$ and a self-propulsion speed $\bfv_p(t)$, which results from
a propulsion force $\bff(t)$ exerted on the fluid. (By Newton third
law, $\bff$ is equal and opposite to $\bfp$ in Eq.~\eqref{eq:dynact}.)
A recording of ${\bfr(t)}$ and ${\bf x}_i(t)$ of duration $t_f$ played
backward is also a solution of Stokes equation.
It would involve a force $-\bff(t_f-t)$, a speed $-\bfv_p(t_f-t)$ and
a flow $-\bfu(t_f-t)$. It is, however, distinguishable from the
forward trajectory: our swimmer will swim backward, a `pusher' would
become a `puller'.
In probabilistic terms, the trajectory $\bfr(t)$ given the displacements ${\bf x}_i(t)$ is equally likely to occur as $\bfr(t_f-t)$ given ${\bf x}_i(t_f-t)$, even though they can be distinguished by the flow they generate.

The situation is reminiscent of the equilibrium Langevin dynamics of a
passive charged particle at ${\bfr}(t)$ in a magnetic field $\bfB(t)$
created by electrons at positions ${\bf x}_i(t)$, moving
deterministically in a coil:
\begin{equation}\label{eq:dynmagfield}
    m\ddot {\bfr} = -\gamma \dot{\bfr}+ q \dot {\bfr} \times \bfB{}(t) +\sqrt{2\gamma k T} \boldsymbol{\eta}\;,
\end{equation}
with $q$ the charge of the particle, $\gamma$ its damping coefficient,
$m$ its mass, and $k T$ the temperature. The reverse trajectories
${\bf x}_i(t_f-t)$ lead to a magnetic field $-\bfB(t_f-t)$ so that
${\bfr}(t_f-t)$ is equally likely to occur as ${\bfr}(t)$ was in the
presence of $\bfB(t)$. Mathematically, the conditional probabilities
of observing $\bfr(t)$ given ${\bf x}_i(t)$ and $\bfr(t_f-t)$ given
${\bf x}_i(t_f-t)$ are thus equal and the corresponding entropy
production rate vanishes $\sigma=\lim_{t\to\infty}\frac 1 t
\log\frac{P[ \{\bfr(t)\} | \{{\bf x}_i(t)\}]} { P [\{\bfr(t_f-t)\} |
    \{{\bf x}_i(t_f-t)\}]}=0$, consistently with thermodynamics.  In a
given experiment, however, it is not the conditional probability which
is measured, but the joint probability of observing ${\bfr}(t)$ and
${\bf x}_i(t)$. When the electrons are driven by a fixed potential
difference, for instance, they will not generate ${\bf x}_i(t)$ and
${\bf x}_i(t_f-t)$ with equal probability. The observation of
${\bfr(t)}$ and ${\bfr(t_f-t)}$ will thus not be equally likely: they
do not inherit the underlying TRS dynamics~\eqref{eq:dynmagfield}
would have if the dynamics of ${\bf x}_i(t)$ were time
reversible. This `induced' irreversibility can be measured by
comparing the occurrence frequency of ${\bfr(t)}$ and ${\bfr(t_f-t)}$
for the same field $\bfB(t)={\bfB_0}$. The corresponding `Shannon'
entropy production rate is now finite, given by $\sigma=2 q^2
|\bfB_0|^2/(\gamma m)$. Note that this `entropy-production rate'
solely measures the irreversibility of the trajectories ${\bfr}(t)$ in
an experiment with a fixed magnetic field. In particular, it is not a
measure of the creation of thermodynamic entropy in our magnetic
system since the magnetic field has to be flipped under
time-reversal. Equation~\eqref{eq:dynmagfield} has also be used to
describe the hair bundle of sensory
cells~\cite{dinis2012fluctuation}. There, the term analogous to
$\bfB(t)$ has a different origin and does not flip under time reversal
and thus leads to a non-vanishing thermodynamic entropy production
rate.  Whether a given definition of $\sigma$ can be connected to the
thermodynamic entropy production rate or not depends on the physics of
the system under study; it can always be connected to a measure of
irreversibility in the sense described above, which is the perspective
adopted in this review.

In active matter, the internal processes leading to ${\bf x}_i(t)$ are
often strongly irreversible. In living systems, they rely on an
imbalance between the concentrations of ATP and
ADP+P~\cite{ajdari1992mouvement,julicher1997modeling} in the
cells. For Janus self-diffusiophoretic colloids, it is the
irreversible transmutation of hydrogen peroxyde into oxygen and
water~\cite{howse2007self,Palacci2010PRL} which powers
self-propulsion. The observation of ${\bfr(t)}$ and ${\bfr(t_f-t)}$
thus need not occur with equal probabilities. (An interesting
exception is when self-propulsion emerges from spontaneous symmetry
breaking as in Quincke rollers~\cite{Bricard:2013:Nature}.) It is this
induced irreversibility that will be discussed in the rest of this
article. Before turning to the corresponding computation of the
entropy production rate, let us stress that the most irreversible
process in active system is, generically, the one generating ${\bf
  x}_i(t)$, and not the dynamics of ${\bfr}(t)$.  When trying to
measure the energy dissipated in an active system, say using
calorimetry, one would expect that this process strongly dominates all
other sources of irreversibility. A large part of the irreversibility
is thus lost if the irreversible process leading to the active force
is not modelled~\cite{pietzonka2017entropy,dadhichi2018origins}. This is, in particular,
the case of Eq.~\eqref{eq:dynact} in which the active force is an
input of the problem whose origin is unspecified. The `dissipation'
measured through $w_p=\langle \bfp \cdot \dot \bfr\rangle$ thus cannot
capture the full irreversibility of the system. As we show below, it
is nevertheless an interesting object of study since it quantifies the
violation of TRS encapsulated in the degrees of freedom ${\bfr}(t)$
and $\bfp(t)$.

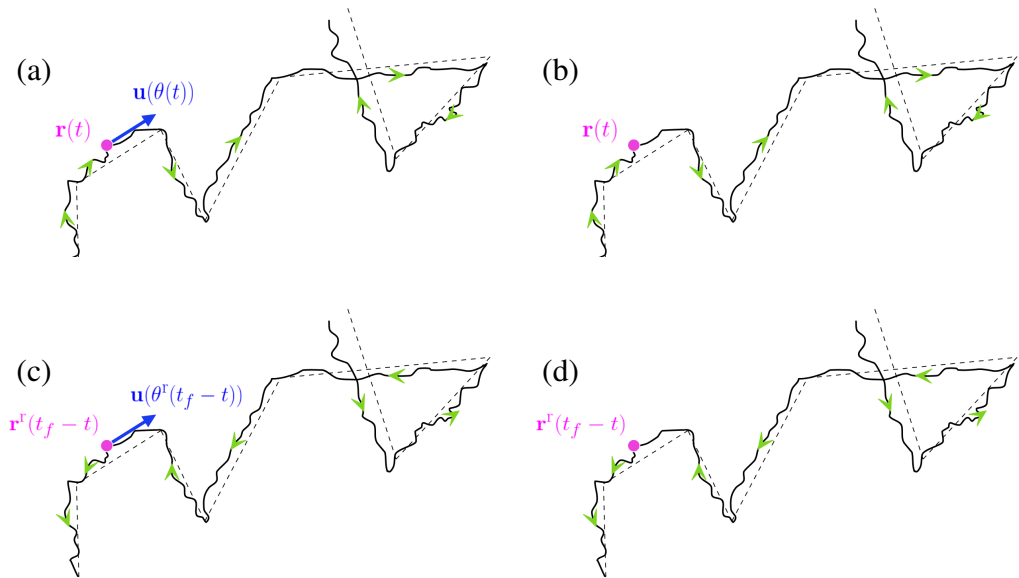
\begin{figure}
\begin{center}
\begin{tikzpicture}
\path (0,0) node {\includegraphics[width=7cm]{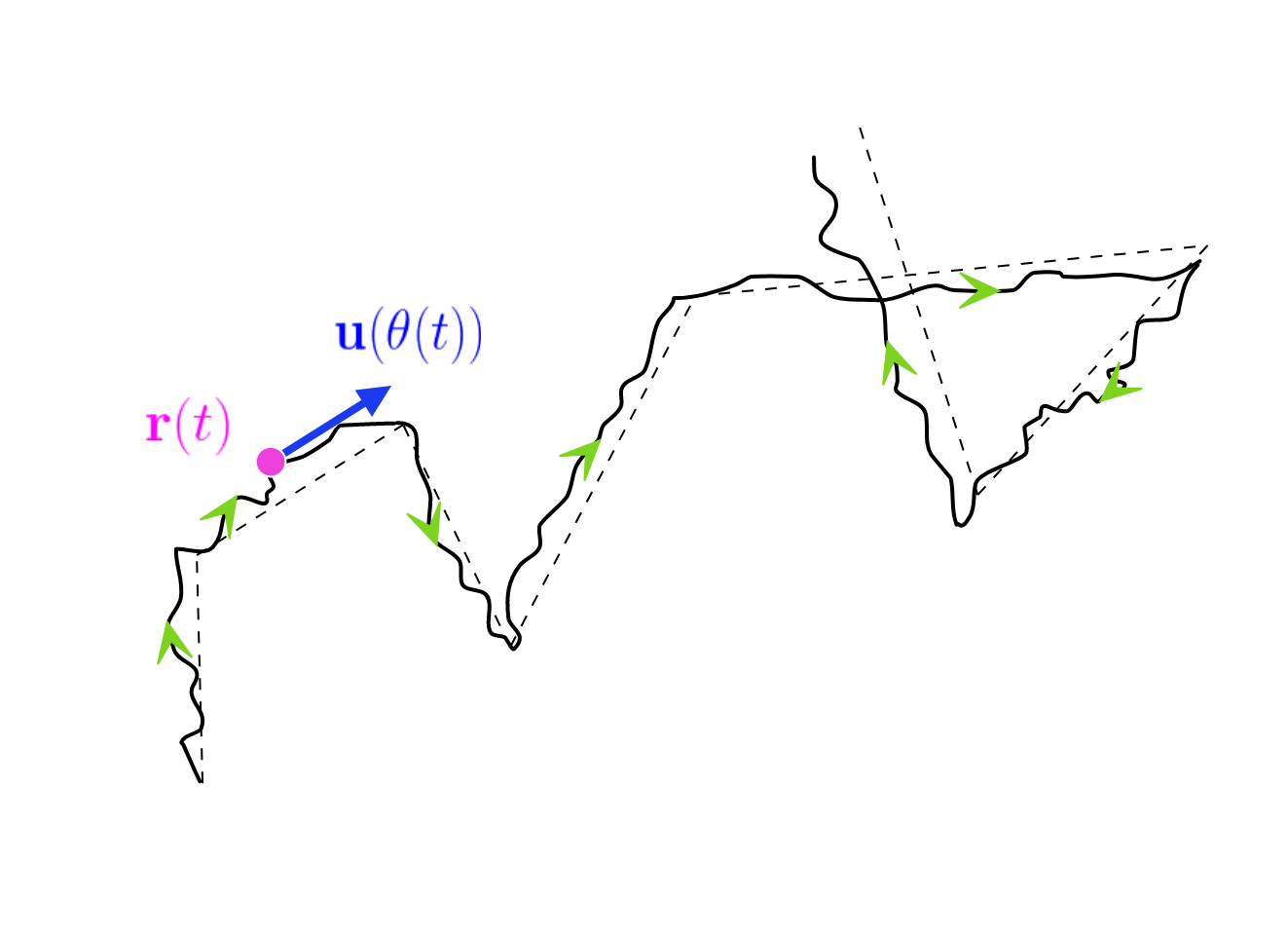}};
\path (0,-4) node {\includegraphics[width=7cm]{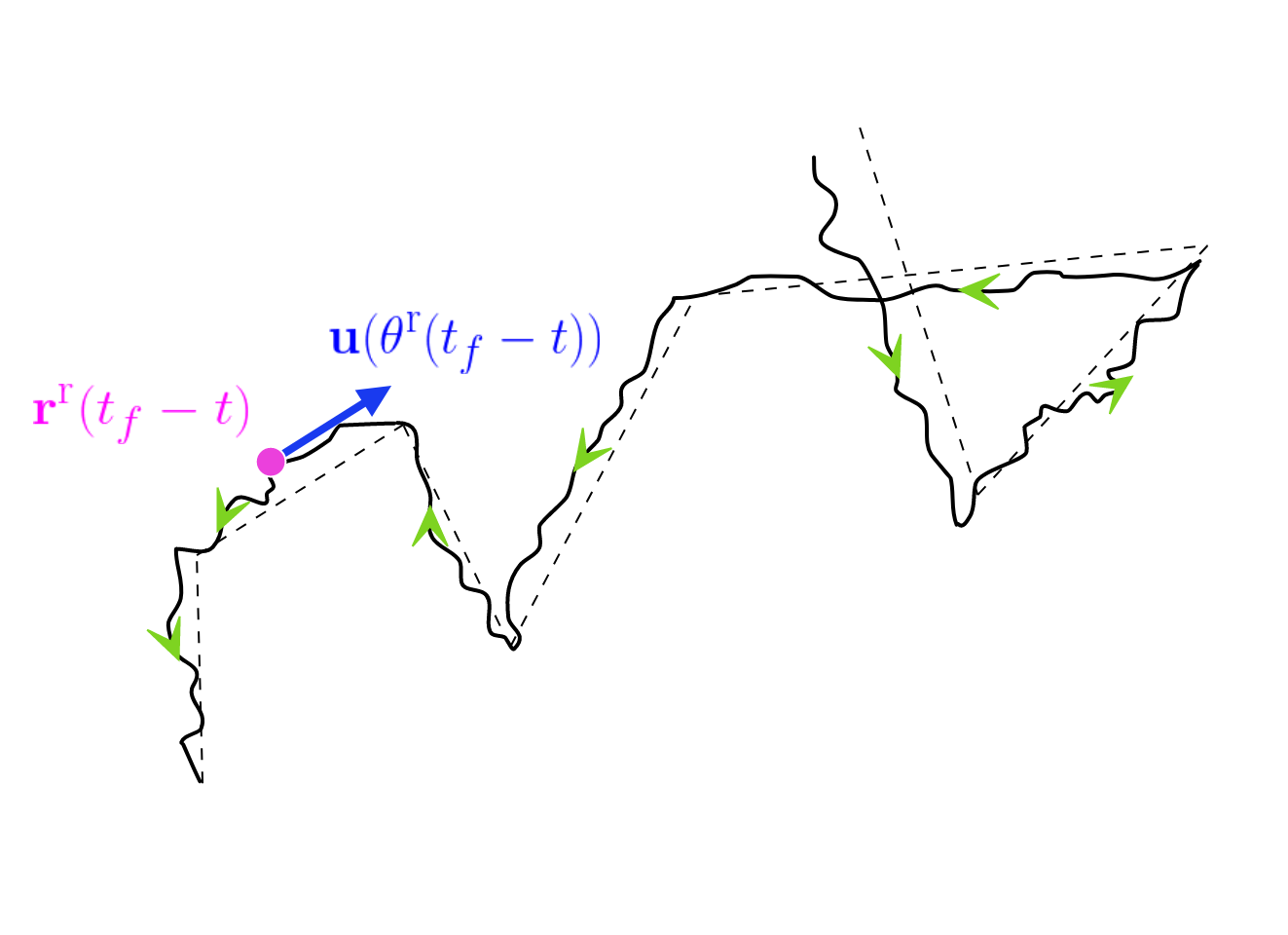}};
\begin{scope}[xshift=7cm]
\path (0,0) node {\includegraphics[width=7cm]{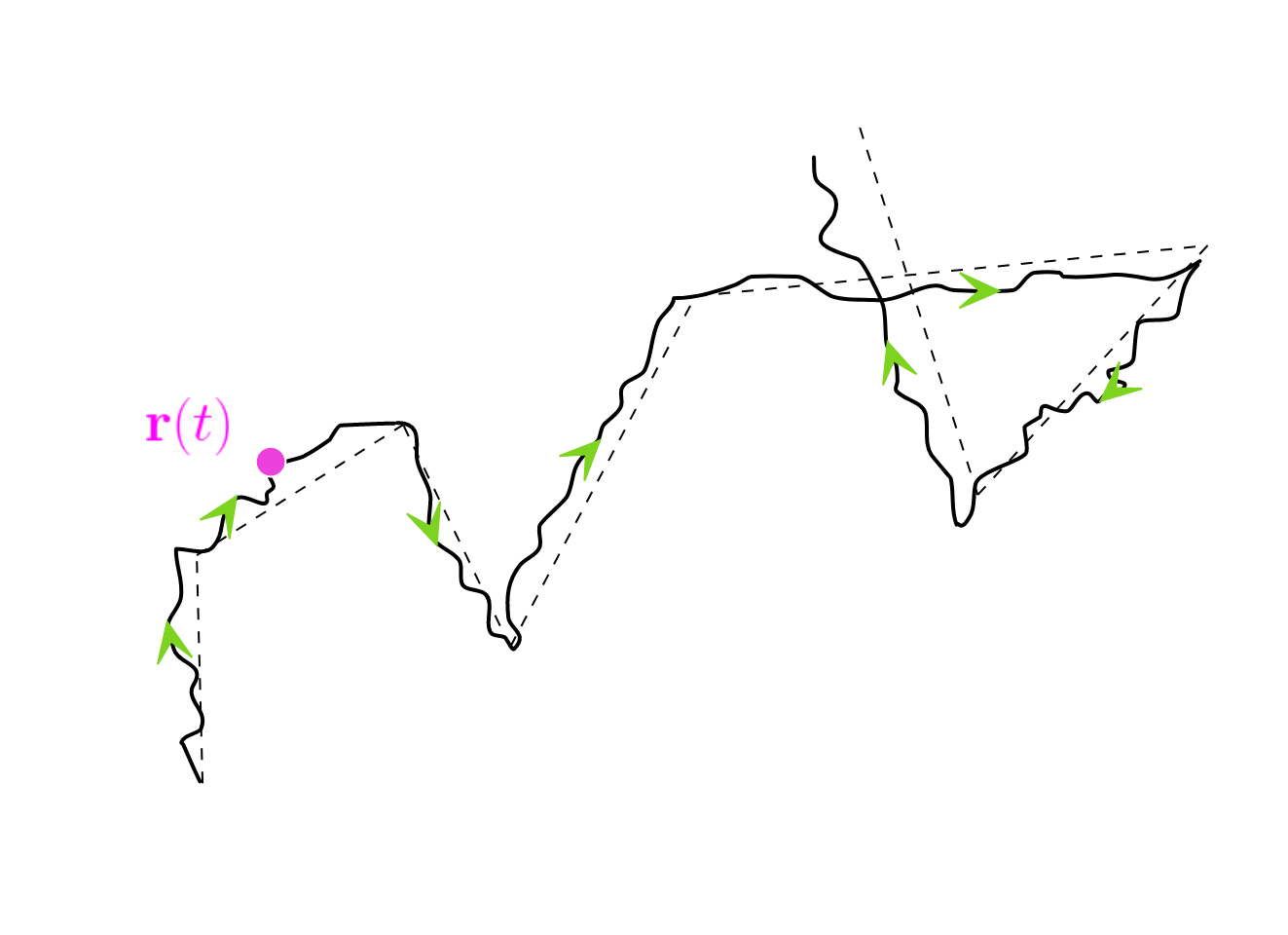}};
\path (0,-4) node {\includegraphics[width=7cm]{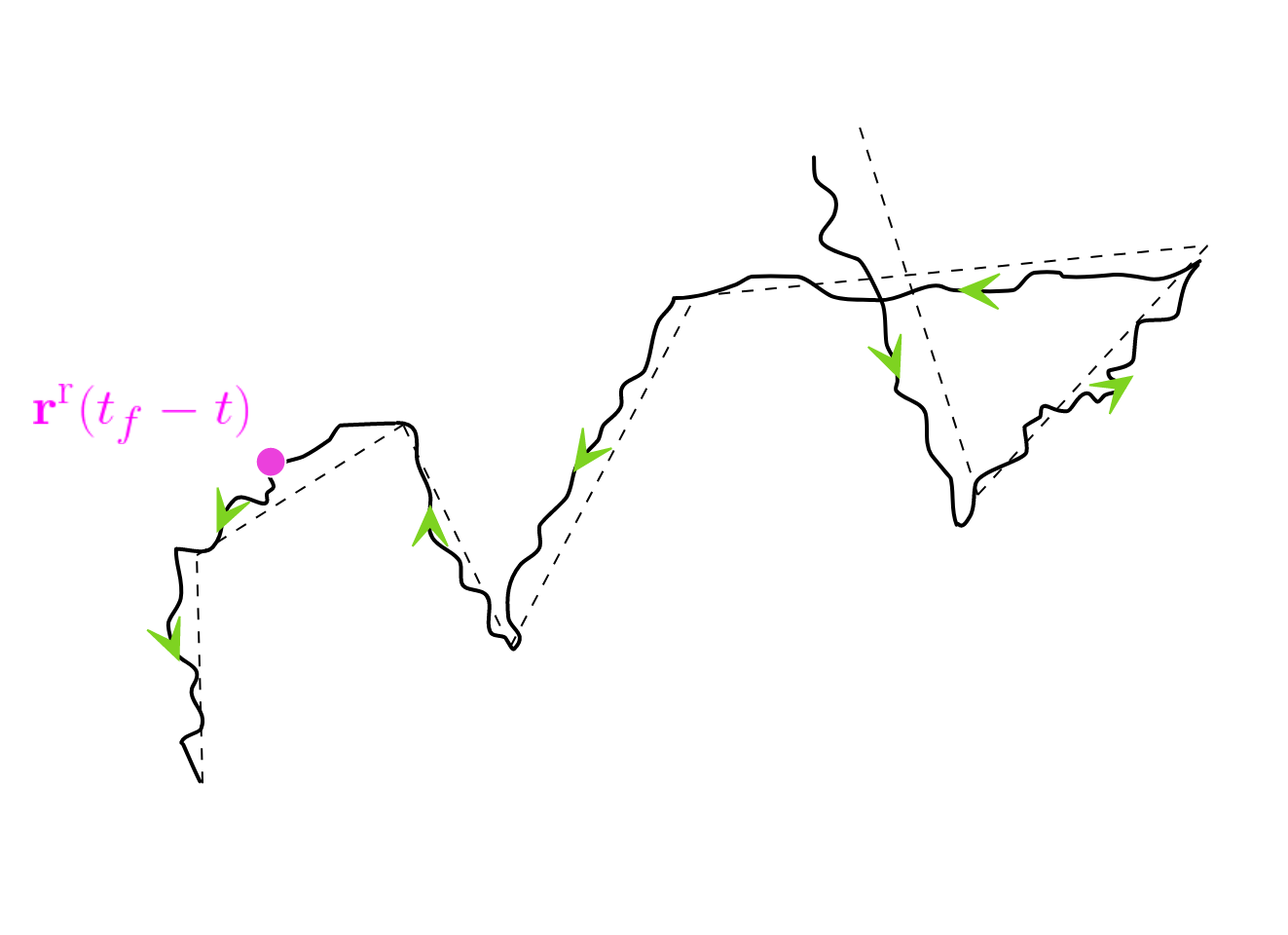}};
\draw (-3,1) node {(b)};
\draw (-3,1-4) node {(d)};
\end{scope}
\draw (-3,1) node {(a)};
\draw (-3,1-4) node {(c)};

\end{tikzpicture}
\end{center}
\caption{Trajectories of a run-and-tumble particle experiencing translational noise. The noise-less trajectory is drawn as a dashed line and the direction of time is indicated by green arrows. One may record the position and orientation of the particle (a) or solely its position (b). The time-reversed trajectories of (a) and (b) are shown in (c) and (d). Their likeliness are clearly different since the noise has to fight self-propulsion in (c), but not necessarily in (d).}\label{fig:TRS}
\end{figure}
Let us first show that, even at the level of
dynamics~\eqref{eq:dynact}, the existence of TRS remains ambiguous and
depends on which degrees of freedom are under study. We start by
considering the situation depicted in Fig.~\ref{fig:TRS}, which
compares the trajectory of a model run-and-tumble bacteria in two
space dimensions with its time-reversed counterpart. The underlying
stochastic dynamics is given by
\begin{equation}\label{eq:dynRTP}
  \dot \bfr(t) = v_0 \bfu(\theta(t)) +\sqrt{2 D} \bfeta(t)
\end{equation}
where $v_0$ is a fixed self-propulsion speed and $\theta(t)$ is fully
randomized at rate $\alpha$. The trajectory presented in
Fig.~\ref{fig:TRS}a records the time evolution of both the particle
position $\bfr(t)$ and its orientation $\theta(t)$.  \if\entroprop1{
  Note that $\bfr(t)$ and $\theta(t)$ uniquely characterize the
  realization of the noise through $ \sqrt{2 D} \bfeta(t)= \dot
  \bfr(t) - v_0 \bfu(\theta(t))$. As detailed in
  Appendix~\ref{app:ep}, for the reversed trajectory $\bfr^{\rm
    r}(t)=\bfr(t_f-t)$, $\theta^{\rm r}(t)=\theta(t_f-t)$ to be
  observed, the surrounding fluid molecules have to produce a
  different noise $\bfeta^{\rm r}(t)$ such that $\sqrt{2D}\bfeta^{\rm
    r}(t)=-\sqrt{2D}\bfeta(t_f-t)-2 v_0 \bfu(\theta(t_f-t))$. This shows that
  time-reversed trajectories are obtained by making the noise
  $\bfeta^{\rm r}$ work against the active force to make the active
  particle retrace its steps.  Using the Gaussian weights of these two
  noise realizations, their relative probability to occur can be
  computed, leading to a path-wise entropy production (See Box
  \textit{Identifying time-reversal symmetry breaking}):
\begin{equation}\label{eq:sigma1bodyhat}
\hat \Sigma [\{\bfr(t),\theta(t)\}] = \frac{\mu}{D} \int_0^{t_f} dt  \dot \bfr \cdot \bfp +\log\frac{P_0(\bfr_0,\theta_0)}{P_f(\bfr_f,\theta_f)},
\end{equation}
where $P_f(\bfr_f,\theta_f)$ is the probability of being at
$\bfr_f\equiv\bfr(t_f)$ and $\theta_f\equiv \theta(t_f)$ given that
the initial condition was sampled according to $P_0$. The right-hand
side of Eq.~\eqref{eq:sigma1bodyhat} measures both the heat
transferred to the bath, $-\hat Q\equiv \int_0^{t_f} \dot \bfr \cdot
\bfp dt $, and the change in the `stochastic' Shannon entropy $\Delta
\hat S=\log \frac{P_0(\bfr_0,\theta_0)}{P_f(\bfr_f,\theta_f)}$ between
$P_0$ and $P_f$ associated to the trajectory $\{\bfr(t),
\theta(t)\}$~\cite{seifert2005entropy}. Taking the average over the
forward path probability, and using the positivity of the
Kullback-Leibler divergence, leads to a generalized second law
$\langle \Delta \hat S \rangle > \frac\mu D \langle \hat Q
\rangle$. Alternatively, taking the limit $t_f\to \infty$, leads to
the steady-state entropy production \textit{rate}
\begin{equation}\label{eq:sigma1bodyfinal}
\sigma = \lim_{t_f\to \infty} \frac{1}{t_f} \hat \Sigma = \frac{\mu \langle \dot \bfr \cdot \bfp \rangle}{D} = \frac{\mu w_p}{D}\;,
\end{equation}
where we have used the ergodicity of the dynamics.}\fi
\if\entroprop2{
  Note that $\bfr(t)$ and $\theta(t)$
uniquely characterize the realization of the noise through
\begin{equation}
  \bfeta(t)=\frac{  \dot \bfr(t) - v_0 \bfu(\theta(t))}{\sqrt{2 D}}
\end{equation}
For the reversed trajectory $\bfr^{\rm r}(t)=\bfr(t_f-t)$,
$\theta^{\rm r}(t)=\theta(t_f-t)$ to be observed, the surrounding fluid
molecules have to produce a different noise
\begin{equation}\label{eq:reversednoise}
  \bfeta^{\rm r}(t)=\frac{  \dot \bfr^{\rm r}(t) - v_0 \bfu(\theta^{\rm r}(t))}{\sqrt{2 D}}=\frac{  -\dot \bfr(t_f-t) - v_0 \bfu(\theta(t_f-t))}{\sqrt{2 D}}=-\bfeta(t_f-t)-\frac{2 v_0 \bfu(\theta(t_f-t))}{\sqrt{2 D}}
\end{equation}
To characterize the irreversibility of
the process it is useful to note that the probability of a trajectory
is given by its Onsager-Machlup
form~\cite{onsager1953fluctuations,woillez:2019:PRL}
\begin{equation}
  P[\{\bfr(t),\theta(t)\}]=Z^{-1} P[\{\theta(t)\}|\theta_0] \exp\Big[-\frac 1 {4D} \int_0^{t_f}dt [\dot \bfr(t)-v_0 \bfu(\theta)]^2\Big] P_0(\bfr_0,\theta_0)
\end{equation}
where $P[\{\theta(t)\}|\theta_0]$ is the probability of the realization of
$\theta(t)$ starting from $\theta_0$, $P_0$ is the probability of the initial condition, and $Z^{-1}$ is a normalization. Note that, for fully
randomizing tumbles, $P[\{\theta(t_f-t)\}|\theta_f]=P[\{\theta(t)\}|\theta_0]$. Taking the logarithm of the ratio between forward and backward trajectories yield the corresponding `path-wise entropy production'~\cite{seifert2005entropy}:
\begin{equation}\label{eq:sigma1body}
\hat \Sigma [\{\bfr(t),\theta(t)\}] \equiv  \log \frac{P[\{\bfr(t),\theta(t)\}]}{P[\{\bfr(t_f-t),\theta(t_f-t)\}]} = \frac{\mu}{D} \int_0^{t_f} dt  \dot \bfr \cdot \bfp +\log\frac{P_0(\bfr_0,\theta_0)}{P_f(\bfr_f,\theta_f)},
\end{equation}
where $P_f(\bfr_f,\theta_f)$ is the probability of being at $\bfr_f\equiv\bfr(t_f)$ and $\theta_f\equiv \theta(t_f)$ given that the initial condition was sampled according to $P_0$. The right-hand side of Eq.~\eqref{eq:sigma1body} measures both the heat transferred to the bath, $-\hat Q\equiv \int_0^{t_f} \dot \bfr \cdot \bfp dt $, and the change in the `stochastic' Shannon entropy $\Delta \hat S=\log \frac{P_0(\bfr_0,\theta_0)}{P_f(\bfr_f,\theta_f)}$ between $P_0$ and $P_f$ associated to the trajectory $\{\bfr(t), \theta(t)\}$~\cite{seifert2005entropy}. Taking the average over the forward path probability lead to a generalized second principle $\langle \Delta \hat S \rangle > \frac\mu D \langle \hat Q \rangle$. Alternatively, taking the limit $t_f\to \infty$, leads to the steady-state entropy production \textit{rate} 
\begin{equation}\label{eq:sigma1bodyfinal}
\sigma = \lim_{t_f\to \infty} \frac{1}{t_f} \hat \Sigma = \frac{\mu \langle \bfr \cdot \bfp \rangle}{D}\;, = \frac{\mu w_p}{D}\;,
\end{equation}
where we have used that the dynamics is ergodic.  }\fi The average
dissipation of the active force $w_p$ thus measures the
irreversibility of the active
dynamics~\eqref{eq:dynRTP}~\cite{Cagnetta:2017:PRL,Nemoto:2019:PRE}. The
latter stems from the velocity being aligned with the self-propulsion
force, hence requiring an atypically strong noise to generate the
time-reversed trajectories. Physically, the entropy production rate
$\sigma$ measures the (inverse) time scale over which the
self-propulsion makes the trajectory irreversible: at short time
scales, the translational diffusion due to Brownian motion dominates
self-propulsion hence hiding the irreversible character of the
dynamics; on longer time scales, translational diffusion plays a
lesser role in transport than self-propulsion, which makes the
irreversibility stemming from the latter more apparent. Note that
$\sigma$ diverges as $D$ goes to zero, so that the dynamics becomes
strongly irreversible in this limit.

The situation is completely different if one tries
to characterize the TRS breaking for the trajectory shown in
Figs.~\ref{fig:TRS}b\&{}d. There, only the position of the particle is
measured and its original and final orientations are unknown. If the
system is endowed with periodic boundary conditions, the steady state
distribution is an isotropic, uniform distribution so that it is
equally likely to find trajectories with $\theta(t)$ or the flipped
orientation $\tilde \theta(t)=\pi-\theta(t)$.
The entropy production thus
vanishes since the time reversed of any trajectory with $\theta(t)$
can be realized with the same probability by a trajectory starting
from the final position $\bfr(t_f)$ with a flipped orientation
$\tilde\theta(t_f)$. Note that this symmetry resembles that of
underdamped Langevin equations in which the time-reversal symmetry in
configuration space emerges from a symmetry in phase-space upon
flipping the velocities under time reversal. This parallel between
active particles and underdamped passive ones has been exploited to
reveal a similar symmetry of the evolution operator of
AOUPs~\cite{Fodor:2016:PRL,mandal2017entropy,caprini2018comment}. Note that this TRS in position space is independent of the value of $D$ and holds even for $D=0$.

Finally, we comment on the fact that whether or not one observes TRS
in active systems depends on what can be measured. Consider for
example Quincke colloidal particles~\cite{Bricard:2013:Nature}. These
particles acquire self-propulsion through the spontaneous breaking of
a symmetry: their polarity stems from an asymmetric charge
distribution on their surface. For an isolated particle, measuring
this asymmetry or the flow of the surrounding fluid are required to
distinguish a forward trajectory from a time-reversed one. In sum, in
a dilute, uniform active system, the observation of a breakdown of TRS
depend on the degrees of freedom which are considered. Describing the
energy source~\cite{pietzonka2017entropy,dadhichi2018origins}, or considering the inertia
of these typically overdamped systems~\cite{shankar2018hidden}, will
lead to different characterization of the irreversibility of the
dynamics. This strongly differs from equilibrium dynamics, in which
TRS holds irrespective of the degree of coarse-graining.

The situation changes drastically when active particles are interacting with their surrounding. While an ambiguity regarding the status of TRS may remain for isolated active particles in the steady state, their interactions with external potentials or with other particles typically reveal their non-equilibrium nature, as we  discuss in
Sections~\ref{sec:obstacles} and~\ref{sec:interactions}. 


\section{Non-interacting active particles in the presence of obstacles and external potentials}
\label{sec:obstacles}
Consider a system in equilibrium in which a small obstacle is
introduced. The corresponding perturbation $\delta V(\bfr)$ leads to a
localized perturbation of the Boltzmann weight $P_{\rm eq}({\bfr})
\propto \exp[-\beta (V(\bfr{})+\delta V(\bfr{}))]$. In that sense, the
perturbation remains local. Furthermore, TRS is maintained so that
even systems with asymmetric obstacles cannot harbor steady currents. Both features
are challenged in active systems, whose fate in the presence of
obstacles and external potentials have attracted a lot of
interest~\cite{angelani2011active,Bechinger:2016:RMP,Baek2018PRL,Cates:2013:EPL,chaudhuri2020active,dhar2019run,elgeti2009self,Enculescu:2011:PRL,fischer2019aggregation,hennes2014self,hermann2018active,koumakis2014directed,Sood:krishnamurthy2016micrometre,kuhr2017collective,malakar2020steady,nash2010run,Nikola2016PRL,Solon2015EPJST,Tailleur:2009:EPL,wagner2017steady}
both for fundamental reasons, in that they probe the relationship
between active and passive dynamics, but also for practical
ones. External potentials and confinements are indeed the toolbox used
to engineer and probe active systems, from optical \& acoustic
tweezers~\cite{takatori2016acoustic} to
centrifuges~\cite{saragosti2011directional} and arrays of
obstacles~\cite{Galajda2007Jbact}.  In addition to a wealth of
experimental
works~\cite{Dileonardo2010PNAS,Sokolov2010PNAS,dauchot2019dynamics,SolonGinot2018NJP,Ginot2015PRX,Palacci2010PRL},
these questions have been addressed at the theoretical level by
considering the overdamped dynamics
\begin{equation}\label{eq:dynVext}
    \dot \bfr = \bfv_p-\mu \nabla V(\bfr)
\end{equation}
where the self-propulsion $\bfv_p$ evolves either through
tumbles~\cite{schnitzer1993theory,Tailleur:2008:PRL,angelani2011active},
rotational diffusion~\cite{Solon2015EPJST,wagner2017steady}, or as an
Ornstein-Uhlenbeck process~\cite{Szamel2014PRE,Fodor:2016:PRL}. In all
cases, one can define a persistence time $\tau$, a typical
self-propulsion speed $v_0$, a persistence length $\ell_p=v_0 \tau$,
and a large-scale diffusivity $D_{\rm eff}\propto \ell_p^2/\tau$.  In
the limit $\tau\to 0$ keeping $D_{\rm eff}$ constant, the dynamics
becomes equivalent to a passive one and leads to a Boltzmann
distribution with an effective temperature $T_{\rm eff}\equiv D_{\rm
  eff}/\mu$. As the persistence time increases, active systems both
develop non-Boltzmann features and exhibit TRS violations. We first
review below the $\tau=0$ limit before discussing the nonequilibrium
static and dynamic features that develop as $\tau$ increases.

\textbf{The $\tau=0$ limit and the universal effective equilibrium
  regime.} For AOUPs, the equilibrium behaviour stems from the fact
that the Gaussian process $\bfv_p$ becomes, as $\tau\to 0$, a white
noise:
$\langle
v_{p,\alpha}(t)v_{p,\beta}(0)\rangle=\delta_{\alpha,\beta}\frac{D}{\tau}
\exp(-\frac{|t-t'|}\tau) \underset{\tau\to 0}{\longrightarrow} 2 D
\delta_{\alpha,\beta} \delta(t-t')$. The dynamics is thus equivalent
to a passive one with a temperature $T_{\rm eff}=D/\mu$. The
effective equilibrium regime also exists for ABPs and RTPs, despite
their non-Gaussian natures. This has been established in any
dimension~\cite{Cates:2013:EPL,Solon2015EPJST} and we detail it here
for RTPs in $d=1$. In the presence of a confining potential $V(x)$,
the steady-state distribution is given
by~\cite{Hanggi1984,Solon2015Nphys} :
\begin{equation}
\label{eq:SSRTPs}
P(x)=\frac{v_0^2 P_0}{v_0^2-\mu^2 V'(x)^2} \exp\left[-\frac\mu\tau 
\int_0^x dx' \frac {V'(x')} { v_0^2-\mu^2 V'(x')^2}\right]\;,
\end{equation}
where $\tau^{-1}$ is the tumbling rate.
Note that Eq.~\eqref{eq:SSRTPs} exhibits, in general, a non-Boltzmann
form: the forces experienced by the particle (that do not stem
from the bath) are not proportional to $\nabla \log P$.
Next, consider the  $\tau\to 0$ limit, keeping $T_{\rm eff}=\frac{v_0^2 \tau}{\mu}$ finite. This implies a large $v_0$ limit, so that $v_0 \gg \mu V'(x)$ for smooth potentials, which allows one to expand~\eqref{eq:SSRTPs} into
\begin{equation}\label{eq:Boltzmanneff}
P(x)=P_0 \exp[-V(x)/T_{\rm eff}]\;.
\end{equation}
\if\selfconsistentpot1{For confining potentials, even if the condition
  $v_0 \gg \mu V'(x)$ cannot hold everywhere, the
  approximation~\eqref{eq:Boltzmanneff} can be shown to be
  self-consistent if $\mu V''(x_0) \tau \ll 1$, where $x_0$ is the
  minimum of the potential, see Appendix~\ref{app:selfcontpot}.  These
  criteria generalize to $v_0 \gg \mu |\nabla V|$ and
  $\mu\tau \Delta V\ll 1$ in higher
  dimensions~\cite{Solon2015EPJST}.}\fi \if\selfconsistentpot2{ Note
  that, for confining potentials, the condition $v_0 \gg \mu V'(x)$
  cannot hold everywhere.  To get a feeling for how small $\tau$
  should be for the approximation to hold, we expand the
  distribution~\eqref{eq:Boltzmanneff} around a minimum $x_0$ of
  $V$. The distribution $P(x)$ is then locally Gaussian, with a
  typical displacement $\sqrt{T_{\rm eff}/V''(x_0)}$.  In turn, the
  typical force scales as
  $V'(x_t)\simeq \sqrt\frac{\tau v_0^2 V''(x_0)}{\mu}$. The condition
  $v_0 \gg \mu V'(x)$ then becomes $\mu V''(x_0) \tau \ll 1$: the
  typical time between two tumbles has to be much shorter than the
  relaxation time inside the potential well.  These criteria
  generalize to $v_0 \gg \mu |\nabla V|$ and $\mu\tau \Delta V\ll 1$
  in higher dimensions~\cite{Solon2015EPJST}.  }\fi

This effective equilibrium regime was demonstrated theoretically for
sedimenting RTPs~\cite{Tailleur:2009:EPL},
ABPs~\cite{Enculescu:2011:PRL,wolff2013sedimentation,Solon2015EPJST,kuhr2017collective,wagner2017steady,hermann2018active,fischer2019aggregation}
and AOUPs~\cite{Szamel2014PRE}, whose sedimentation profiles have been
computed theoretically and lead to Eq.~\eqref{eq:Boltzmanneff} in the
small $\tau$ limit. Experimentally, the effective equilibrium regime
has been measured for sedimenting self-propelled diffusiophoretic
colloids~\cite{Palacci2010PRL,Ginot2015PRX,SolonGinot2018NJP}. For
ABPs, RTPs and AOUPs in harmonic traps, the effective equilibrium
regimes have also been studied
theoretically~\cite{Tailleur:2009:EPL,Szamel2014PRE,takatori2016acoustic,malakar2020steady}
and measured in experiments~\cite{takatori2016acoustic}. Probing
effective equilibrium regimes in more general experimental settings
remains an open challenge~\cite{han2017effective}.

\textbf{Departure from the $\tau=0$ limit: non-Boltzmann
  distributions.} Non-thermal effects have naturally been the focus of
the community and the departure from the $\tau=0$ limit is
particularly relevant from that perspective. Despite a universal
$\tau=0$ regime, different models of self-propelled particles have
been shown to lead to different behaviours, both from a static and a
dynamic perspective, as soon as $\tau\neq 0$. General expressions for
arbitrary potentials have been obtained for a single RTP in one
dimension to any order in $\tau$, see Eq.~\eqref{eq:SSRTPs}. For
AOUPs, many different approaches have been developed. Some are based
on calculating the steady-state directly using either path
integrals~\cite{bray1990path,mckane1990path,woillez2020nonlocal} or
perturbative
approaches~\cite{klosek1988colored,Fodor:2016:PRL,bonilla2019active,martin2020statistical,martin2020aoup}. Others
rely on effective equilibrium approximations of the
dynamics~\cite{Fox:86a,Fox:86b,Hanggi:87,Cao:93,maggi2015multidimensional,wittmann2017effective1,wittmann2017effective2,farage2015effective,marconi2015towards}. As
a result, the steady-state distribution for $N$ interacting AOUPs has
been obtained to order $\tau$ in any
dimension~\cite{maggi2015multidimensional,Fodor:2016:PRL,bonilla2019active,martin2020statistical}. For
a single AOUP, it has been obtained explicitly up to order $\tau^2$
using a perturbative expansion that can be extended to higher
orders~\cite{Fodor:2016:PRL,bonilla2019active,martin2020statistical,martin2020aoup}. We
use these results below to illustrate and contrast the departure of
the steady-state distribution from its $\tau=0$ limit for both RTPs
and AOUPs.

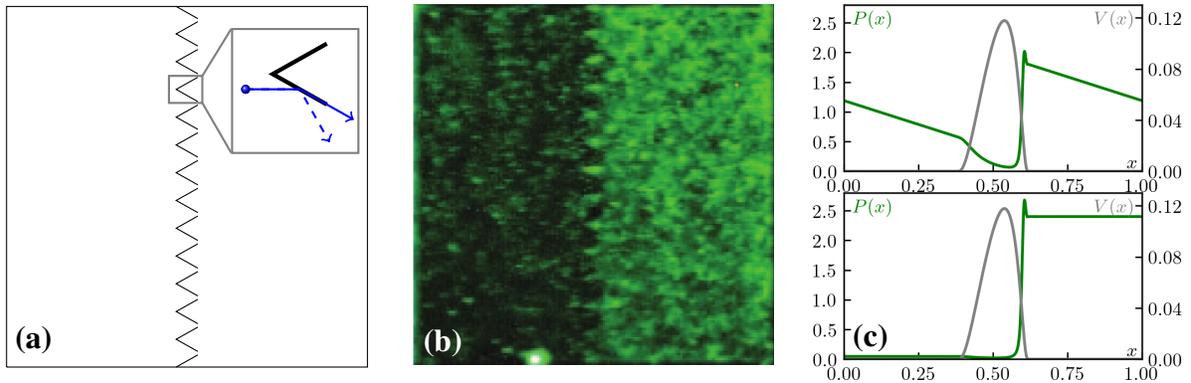
\begin{figure}
  \centering
  \hspace{0.5cm}\begin{tikzpicture}[scale=.012]
    \draw (-170,-170) node {\bf (a)};
    \draw (-200,-200) -- (-200,200) -- (200,200) -- (200,-200) -- (-200,-200);
    \draw (-12,-200) -- (12,-200+13.5);
    \draw (12,-200+13.5+3.8) -- (-12,-200+27+3.8) -- (12,-200+27+13.5+3.8);
    \draw (12,-200+27+13.5+2*3.8) -- (-12,-200+2*27+2*3.8) -- (12,-200+2*27+13.5+2*3.8);
    \draw (12,-200+2*27+13.5+3*3.8) -- (-12,-200+3*27+3*3.8) -- (12,-200+3*27+13.5+3*3.8);
        \draw (12,-200+3*27+13.5+4*3.8) -- (-12,-200+4*27+4*3.8) -- (12,-200+4*27+13.5+4*3.8);
        \draw (12,-200+4*27+13.5+5*3.8) -- (-12,-200+5*27+5*3.8) -- (12,-200+5*27+13.5+5*3.8);
        \draw (12,-200+5*27+13.5+6*3.8) -- (-12,-200+6*27+6*3.8) -- (12,-200+6*27+13.5+6*3.8);
        \draw (12,-200+6*27+13.5+7*3.8) -- (-12,-200+7*27+7*3.8) -- (12,-200+7*27+13.5+7*3.8);
        \draw (12,-200+7*27+13.5+8*3.8) -- (-12,-200+8*27+8*3.8) -- (12,-200+8*27+13.5+8*3.8);
        \draw (12,-200+8*27+13.5+9*3.8) -- (-12,-200+9*27+9*3.8) -- (12,-200+9*27+13.5+9*3.8);
        \draw (12,-200+9*27+13.5+10*3.8) -- (-12,-200+10*27+10*3.8) -- (12,-200+10*27+13.5+10*3.8);
        \draw (12,-200+10*27+13.5+11*3.8) -- (-12,-200+11*27+11*3.8) -- (12,-200+11*27+13.5+11*3.8);
        \draw (12,-200+11*27+13.5+12*3.8) -- (-12,-200+12*27+12*3.8) -- (12,-200+12*27+13.5+12*3.8);
        \draw (-12,200) -- (12,200-13.5);
        \draw[thick,gray] (-20,93) rectangle +(37,30);
        \draw[gray,thick] (17,93)--(50cm-30,50cm-350);
        \draw[gray,thick] (17,123)--(50cm-30,50cm+125cm);
        \begin{scope}[scale=2.5,xshift=50cm,yshift=50cm]
        \draw[thick,gray] (-30,-35) rectangle +(56,55);
                  \draw[ultra thick] (12,-13.5) -- (-12,0) -- (12,13.5);
          \draw[blue,thick,->] (-24,-6.75) -- (0,-6.75) -- (24,-20.25);
          \shade[ball color=blue] (-24,-6.75) circle (2);
          \draw[blue,dashed,thick,->] (-24,-6.75) -- (0,-6.75) -- (13.24,-6.75-23.53);
          \shade[ball color=blue] (-24,-6.75) circle (2);
      \end{scope}
    \end{tikzpicture}
    \vspace{-.25cm}
    \hspace{0.25cm}
    \raisebox{-.1cm}{\begin{tikzpicture}
          \node (0,-0.2) {\includegraphics[width=.29\textwidth,totalheight=.29\textwidth]{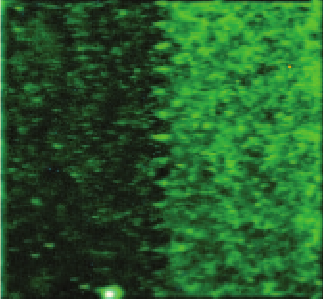}};
          \draw[white] (-2,-2.1) node {\bf (b)};
        \end{tikzpicture}}
      \raisebox{-0.3cm}{
        \begin{tikzpicture}
        \path (0,0) node {\includegraphics[width=5cm]{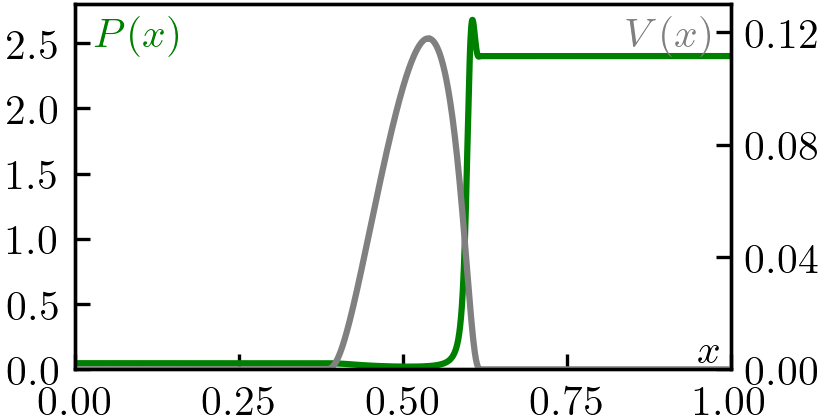}};
        \path (0,2.5) node {\includegraphics[width=5cm]{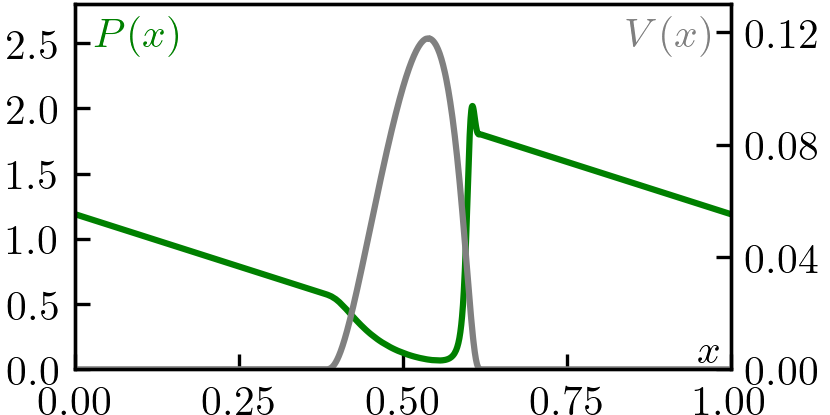}};
        \draw (-1.7,-0.7) node {\bf (c)};

      \end{tikzpicture}}
    \caption{{\bf (a) \& (b) } Microfluidic chamber in which
      run-and-tumble bacteria or colloidal particles can be
      inserted. An array of asymmetric obstacles split the cavity into
      two regions {\bf (a)}. While colloidal particles would lead to a
      uniform density away from the obstacles, bacteria accumulate in
      one side of the system {\bf (b)}, as shown by their fluorescence
      signal~\cite{Galajda2007Jbact}. {\bf Inset of (a):} The
      non-Boltzmann distribution solely stems from the interaction
      between particles and walls. Replacing the aligning torques
      experienced by the bacteria upon encountering a wall (solid blue
      line) by a specular reflection (dashed blue line) of their
      orientation would lead to a time-reversal symmetric dynamics and
      a uniform density~\cite{Tailleur:2009:EPL}. The arrow of time
      can be read in the bacterial dynamics (solid blue line) by
      comparing the occurence probability of forward and backward
      trajectories. The former requires no tumble to occur during a
      time $\tau$ whereas the latter requires a tumble to return to
      the original position. {\bf (c)} Steady-state density profiles
      (green) generated by an asymmetric obstacle (grey) placed at the
      center of the system. Closed boundary conditions (bottom) lead
      to a density difference in the two compartments whereas periodic
      boundary conditions (top) lead to a linear profile away from the
      obstacle and a non-zero current. }
    \label{fig:austin}  
\end{figure}

In both cases, the steady-state distribution in the presence of a confining potential $V(x)$ can be written, in one dimension, as in Eq.~\eqref{eq:Boltzmanneff}, albeit with $V(x)$ replaced by an effective potential $V_{\rm eff}(x)$, which can be computed perturbatively.
For an AOUP, one finds 
\begin{eqnarray}\label{eq:AOUPST}
    V_{\rm eff}(x)&=&V(x)- \tau \left( T_{\rm eff}{V''(x)}-\frac{V'(x)^2}2\right)\\
  &&\!\!\!\!\!\!\!\!-\tau^2 \left(\frac{T_{\rm eff}^2 V^{(4)}(x)}2 + \frac{\int^x V'(y)^2 V^{(3)}(y)dy}2 - T_{\rm eff} V'(x) V^{(3)}(x)- T_{\rm eff}\frac{V''(x)}4 \right) +{\cal O}(\tau^3)\;.\notag
\end{eqnarray}
A number of interesting features can already be noted in this
perturbative expansion. First, a purely repulsive potential $V(x)$ may
lead to an effective potential with attractive parts, due to the term
$-\tau T_{\rm eff} V''(x)$. While derived perturbatively, this gives a
heuristic explanation for the accumulation of active particles close
to walls, which is a trademark of active
particles~\cite{elgeti2009self,wagner2017steady,yang2014aggregation,ezhilan2015distribution,elgeti2013wall,sartori2018wall}. Second,
an important difference between passive and active systems can be
observed at order $\tau^2$ in Eq.~\eqref{eq:AOUPST}: The steady-state
distribution $P(x)$ is a non-local functional of $V$ for active
particles. Consider a dilute system. In thermal equilibrium, a
perturbation of the potential $\delta V$ localized at $y$ does not
impact
$P(x\neq y)\propto e^{-\beta [V(x)+\delta V(x)]}=e^{-\beta V(x)}$, up
to an overall normalization. In the active case, Eq.~\eqref{eq:AOUPST}
reveals a completely different behaviour:
\if\alt1{$P(x \neq y)$ now depends on $V(y)$ for arbitrary large
  values of $|x-y|$, as exemplified in Appendix~\ref{app:nonlocal}.}\fi{}
\if\alt2{$P(x \neq y)$ now depends on $V(y)$ for arbitrary large
  values of $|x-y|$. To see this consider the perturbation
  $\delta V(x)=\epsilon \delta(x-y)$. To linear order in $\epsilon$,
  the effective potential at $x$ picks up a contribution
  $\delta V_{\rm eff}(x)= \frac{\epsilon \tau^2}{2}
  [(V'(y)^2)'''+2(V'(y)V'''(y))']\Theta(x-y)$, which adds a global
  step to the density profile at $x=y$.}\fi{}
This is a simple heuristic explanation of remarkable experiments
conducted on swimming bacteria that show an array of asymmetric
obstacles to act as a pump when placed in the middle of a microfluidic
cavity~\cite{Galajda2007Jbact} (See Fig.~\ref{fig:austin}).

The derivation of the steady-state distribution of a single RTP can also be carried out using Eq.~\eqref{eq:SSRTPs}, yielding
\begin{equation}\label{eq:RTPVeff}
 V_{\rm eff}(x)=V(x)-\frac{\mu \tau}{T_{\rm eff}} \left( (V'(x))^2+\frac{1}{T_{\rm eff}} \int^x dy  (V'(y))^3 \right)+{\cal O}(\tau^2)\;.
\end{equation}
Again, both the emergence of effective attractive interactions out of repulsive potentials and the non-locality of $P(x)$ emerge as $\tau$ departs from $0$. Contrary to AOUPs, however, both effects are already present at order $\tau$, hence highlighting the non-universality of the departure from the $\tau=0$ equilibrium limit across models. 

Much less is known in higher dimensions, where exact results and
controlled perturbative expansions are harder to get. The far-field
perturbations due to localized objects have been
characterized~\cite{Baek2018PRL} and the steady-state distributions of
AOUPs can be obtained
perturbatively~\cite{Fox:86a,Fox:86b,Hanggi:87,Cao:93,maggi2015multidimensional,wittmann2017effective1,wittmann2017effective2,farage2015effective,marconi2015towards,Fodor:2016:PRL,martin2020statistical}. Interestingly,
both the effective attractions and the non-local corrections to the
density field remain present. The perturbation to the density field
induced by an asymmetric obstacle now decays as a dimension-dependent
power-law of the distance to the object~\cite{Baek2018PRL}. These
results on the impact of an asymmetric object have also been extended
to active particles experiencing pairwise
interactions~\cite{granek2020bodies}. By contrast, in the passive
case, a perturbation of the density field on the scale of the
correlation length is expected. Away from criticality,
these are short-range effects and hence much weaker than those found
in active systems. Finally, these power-law corrections to the density
field have striking consequences in the presence of a disordered
potential, as examplified by the suppression of the motility-induced phase
separation~\cite{ro2020disorder}.

To illustre the discussion above, we now come back to the examples of
sedimenting and trapped active particles whose effective equilibrium
regimes were discussed before. In the case of sedimenting active
particles, the sedimentation profiles remain given by the Boltzmann
weight in their distal region even when the condition
$v_0 \gg \mu \nabla V$ does not hold:
$\rho(z)\propto \exp(-\delta m g z/\lambda)$ where $\delta m g$ is the
effective weight of the particles and $\lambda$ is a system-dependent
constant that differs for AOUPs, ABPs and RTPs. For AOUPs, $\lambda$
is always equal to $T_{\rm eff}$~\cite{Szamel2014PRE}, even outside
the small $\tau$ regime.  By contrast, $\lambda$ explicitly depends on
$\delta m g$ for ABPs and RTPs, and leads to a gravitational collapse,
$\lambda=0$, when the sedimentation speed $\mu \nabla V$ equals the
self-propulsion one~\cite{Tailleur:2009:EPL}. Next, we turn to the
harmonic confinement of active
particles~\cite{Tailleur:2009:EPL,nash2010run,pototsky2012active,hennes2014self,Solon2015EPJST,takatori2016acoustic,dauchot2019dynamics,dhar2019run,basu2020exact,chaudhuri2020active,malakar2020steady}. First,
results on AOUPs suggest that, as for sedimentation, the equilibrium
phenomenology survives: the steady state remains a Gaussian, albeit
with a potential-dependent effective temperature~\cite{Szamel2014PRE}
and the entropy production vanishes~\cite{Fodor:2016:PRL}. This is in
stark contrast with ABPs and RTPs for which a trapping force
$F({\bfr})=-k \bfr$ leads to a finite horizon
$r_{\rm max}=v_0/(\mu k)$. The Gaussian behaviour in the small $\tau$
limit is replaced by a sharp density accumulation at $r=r_{\rm max}$
in the opposite
limit~\cite{Tailleur:2009:EPL,Solon2015EPJST,takatori2016acoustic,dhar2019run,basu2020exact,chaudhuri2020active}. It
is interesting to note that AOUPs, which are praised for their
simplicity, miss, in this case, an important feature of active
dynamics.

\noindent\textbf{Departure from the $\tau=0$ limit: the emergence of
  non-equilibrium dynamical features.}  As discussed in the Box
\textit{Identifying time-reversal symmetry breaking}, the departure
from thermal equilibrium is not only signaled by a non-Boltzmann
distribution but, most importantly, by the emergence of novel
dynamical features resulting from the breakdown of TRS. This can
already be seen at the non-interacting level in the presence of
external potentials for large enough persistence times. Some of these
nonequilibrium dynamical features have recently attracted a lot of
attention and are discussed below.

\begin{enumerate}
\item \textbf{Steady-state currents.} One of the most important
  results in non-equilibrium statistical mechanics is the generic
  emergence of steady-state currents resulting from the interplay
  between the breakdown of TRS and the lack of spatial symmetries. From
  Brownian
  ratchets~\cite{smoluchowski1927experimentell,feynman1965feynman,magnasco1993forced,parrondo1996criticism,sekimoto1997kinetic,hanggi2009artificial}
  to molecular motors~\cite{ajdari1992mouvement,julicher1997modeling},
  this has been exemplified first by considering non-equilibrium
  dynamics in the presence of asymmetric potentials. These results
  extend to active particles, whose isotropic motions are also
  rectified in the presence of asymmetric potentials, leading to the
  emergence of steady-state
  currents~\cite{ai2013rectification,pototsky2013rectification,koumakis2014directed,ghosh2013self,yariv2014ratcheting,stenhammar2016light}.

Analytically, the current has been computed for AOUPs~\cite{martin2020statistical,martin2020aoup} and RTPs~\cite{angelani2011active} in one dimension in the presence of a periodic, asymmetric potential. Perturbatively in $\tau$, the currents scale as $\tau^2$ and $\tau$ for AOUPs and RTPs, respectively. This highlights, once again, the difference between these models as they depart from the effective equilibrium regime. Mathematically, it can be traced back to the order in $\tau$ at which the non-local terms appear in Eqs~\eqref{eq:AOUPST} and~\eqref{eq:RTPVeff}.

These results can easily be generalized to a single localized obstacle
in the presence of periodic boundary conditions. In one space
dimension, the current then decays as the inverse system size and a
linear density profile is observed away from the obstacle (See
Fig~\ref{fig:austin}(c)).  These results generalize to higher
dimensions where asymmetric obstacles also generate currents. In the
case of localized obstacles, the currents are long ranged and decay
with the distance $r$ to the object as a dimension-dependent
power-law~\cite{Baek2018PRL,granek2020bodies}: $|{\bf J}|\propto
r^{-d}$.

The presence of a current signals a non-zero mean force exerted by the
obstacle on the active fluid. By Newton's third law, a current implies
a net force exerted by the fluid on the object. This explains, for
instance, why an asymmetric gear immersed in a bacterial bath exhibits
a persistent biased rotating motion in the steady
state~\cite{Sokolov2010PNAS,Dileonardo2010PNAS,maggi2015micromotors}. From
a mechanical balance perspective, relevant in these overdamped
systems, a non-zero mean current implies a non-zero drag, which has to
be balanced by a net force from the object, whence relating
currents and  forces in active
systems~\cite{Nikola2016PRL,granek2020bodies}.

The generic emergence of currents outside equilibrium when spatial
symmetries are broken displays interesting exceptions. In active
systems, for instance, currents are absent if, instead of using
asymmetric potentials, one enforces an asymmetric spatial modulation
of the self-propulsion speed~\cite{stenhammar2016light}, despite
non-homogeneous steady-state density
profiles~\cite{schnitzer1993theory,Tailleur:2008:PRL,arlt2018painting,arlt2019dynamics,frangipane2018dynamic}. The
currents are, somewhat counter-intuitively, restored upon the addition
of pairwise forces~\cite{stenhammar2016light}. This is reminiscent of
the physics of passive particles in the presence of an asymmetric
modulation of the temperature field where currents are only observed
in the presence of interactions between 
particles~\cite{VK1988IBM}.

\item \textbf{Entropy production.} The computation of the entropy
  production rate for a non-interacting run-and-tumble particle
  presented in section~\ref{subsec:TRSnointeraction} can be
  generalized to the case of an external force $\bfF$: $\dot\bfr = \mu
  \bfp + \mu \bfF +\sqrt{2 D} \boldsymbol{\eta}$. When computing the
  entropy production in the full $(\bfr,\bfp)$ space, one then finds $
  D\sigma/\mu=\langle \dot\bfr \cdot \bfp\rangle+\langle \dot\bfr
  \cdot {\bf F}\rangle $. For a conservative force $\bfF=-\nabla
  V(\bfr)$, this reduces to $\sigma=\frac{\mu}D \langle \dot\bfr \cdot
  \bfp\rangle =\frac{\mu}D w_p$. The absence of $V$ in this second
  formula does not imply an entropy production rate independent of the
  potential: the dissipation $w_p$ indeed depends on $V(\bfr)$ through
  the precise form of the steady-state distribution.

Integrating out $\bfp$ to compute the entropy production rate $\tilde
\sigma$ in $\bfr$-space is, however, not as easy as in the absence of
external force and there is no general expression for $\tilde
\sigma$. We simply know that coarse-graining these mesoscopic
stochastic dynamics can only lower their entropy production rate so
that $\tilde \sigma \leq \sigma$ (see,
e.g.,~\cite{roldan2018arrow}). Progress has been made when the active
noise is a Gaussian process, as is for instance the case for
AOUPs. There, the computation has been carried out for $N$ interacting
particles in $d$ dimensions, in the presence of forces stemming from a
potential $V$, leading
to~\cite{Fodor:2016:PRL,grafke2017spatiotemporal,martin2020statistical,martin2020aoup}
$\sigma=\frac{\tau^2} {2 D } \langle (\dot \bfr \cdot \nabla)^3 V(\bfr)
\rangle$. This expression simplifies in the small persistence time
limit to give $\sigma= \frac{D \tau^2}2 \langle \sum_{i,j,k}
(\nabla_{i} \nabla_{j} \nabla_{k} V)^2 \rangle $. It is remarkable
that, to order $\tau$, AOUPs are time-reversal symmetric, despite
their steady-state distribution already being different from the
Boltzmann weight.

\item \textbf{Activated events, non-Arhenius law and Onsager-Machlup symmetry.} In equilibrium, an important consequence of TRS is the celebrated Onsager-Machlup symmetry~\cite{onsager1953fluctuations,machlup1953fluctuations} which states that, in a macroscopic system (or in a small-noise limit), the most likely path to realize a rare excursion follows the time-reversal of the most likely relaxation from this rare event into the most probable state. This has important consequences, for instance, for reaction-rate theory where it implies that, in the small noise limit, transitions between states are realized by following steepest-gradient ascent and descent in a free energy landscape. This symmetry breaks down out of equilibrium, an effect that has attracted a lot of interest~\cite{bertini2002macroscopic,tailleur2008mapping,bertini2015macroscopic} and may lead to non-equilibrium dynamical phase transitions~\cite{bodineau2005distribution,bodineau2008long,bunin2013cusp,baek2015singularities}. 

Apart from recent results on nucleation pathways in
MIPS~\cite{richard2016nucleation,redner2016classical,levis2017active},
reaction-rate theory remains largely uncharted territory for active
  matter. Interesting differences with equilibrium dynamics have
  already been observed at the level of first-passage time
  computations~\cite{bray1990path,angelani2014first,woillez2020nonlocal}. In
  the low-noise regime, Arhenius law is, for instance, not
  valid~\cite{woillez:2019:PRL} and the barrier crossing is controlled
  by a combination of force balance and potential differences.

\item \textbf{Linear response} studies the reaction of systems to
  small perturbations. For Markov processes, even out of equilibrium,
  the Agarwal formula generically predicts the response of an
  observable $A$ to the perturbation of the system by a field $B$ from
  the knowledge of steady-state
  correlations~\cite{agarwal1972fluctuation,prost2009generalized,baiesi2013update}. This
  formula also requires the knowledge of the steady-state distribution
  and it only reduces to the celebrated fluctuation-dissipation
  theorem (FDT) when the steady state is given by a Boltzmann
  weight. Experimentally, the FDT allows the characterization, at the
  microscopic scale, of passive systems using microrheological,
  mechanical measurements.

  A first approach to linear response in active systems relied on
  measurements of `violations' of the equilibrium FDT, or lack
  thereof. This first allows identifying regimes in which an
  `effective temperature' can be
  defined~\cite{loi2008effective,morozov2010motor,loi2011non,Wang2011JCP,loi2011effective,bohec2013probing,suma2014dynamics,Szamel2014PRE,levis2015single,Solon2015EPJST,han2017effective,cugliandolo2019effective}. Then,
  the breakdown of FDT can also be used to identify the time scales
  over which activity drives biological systems out of equilibrium,
  both in vitro~\cite{Mizuno2007Science} and in
  vivo~\cite{wilhelm2008out,robert2010vivo,guo2014probing,fodor2015activity,fodor2016nonequilibrium,fodor2018spatial,bohec2019distribution}.
An alternative approach is to start from a microscopic model of active systems and explicitly construct its linear response~\cite{maes2020fluctuating}. This has, in particular, been carried out using perturbative expansions~\cite{Fodor:2016:PRL,martin2020statistical,cengio2020fluctuation} or Markovian approximations~\cite{dal2019linear,caprini2018linear} and might pave the way towards a systematic microrheology of active systems.

\end{enumerate}

\section{TRS in the presence of interactions in active systems}
\label{sec:interactions}

One of the most striking features of active systems is the wealth of dynamical phenomena it exhibits that are without counterparts in equilibrium. From the emergence of travelling waves in colloidal rollers~\cite{Bricard:2013:Nature} to the collective migration of cell monolayers~\cite{poujade2007collective}, a trademark of active systems is the emergence of macroscopic steady-state currents (See Fig.~\ref{fig:collective}). This is a clear TRS-breaking, non-equilibrium feature that allows to immediately distinguish active from passive systems in many situations. Mechanistically, as exemplified by the physics of flocking~\cite{Vicsek:1995:PRL,Toner:1995:PRL}, 
this naturally emerges in phase transitions whose order parameters are coupled to the particle orientations. Estimating the entropy production rate in these strongly irreversible systems is an open problem, on which progress has been made recently using field-theoretical descriptions of flocking models~\cite{borthne2020time}. Note that, while the global phenomenology of a model can be strongly out of equilibrium some of its features may retain an equilibrium nature, as was for instance suggested for angular correlations between birds in starling flocks~\cite{Mora2016NatPhys}. Identifying the degrees of freedom that contribute to TRS violations is thus also an open challenge~\cite{Battle2016Science,gladrow2016broken,Nardini:PhysRevX.7.021007,martin2020statistical}.

Microscopically, the breakdown of TRS emerges from very simple
interactions between particles, even without relying on
alignment. Indeed, mutual hard-core exclusion suffices to make active
dynamics irreversible~\cite{mallmin2019exact} (See Box
\textit{Violation of TRS resulting from hard-core interactions} for a
simple example). A natural question is then whether TRS survives in
some classes of interacting active-matter systems. We discuss below
two such instances. First, we consider MIPS in which TRS is partially
restored by coarse-graining in certain classes of systems. We then
discuss under which conditions coarse-graining may restore TRS
exactly, considering in particular tactic systems. Finally, we turn to
the small-persistence-time regime in which an exact TRS is recovered
for interacting AOUPs despite a non-Boltzmann distribution.

\begin{tcolorbox}[breakable]
\textbf{Violation of TRS resulting from hard-core interactions.}
Consider the following one-dimensional lattice model of run-and-tumble particles~\cite{Thompson:2011:JSM,Soto:2014:PRE,mallmin2019exact}: \pluspart{} and \minuspart{}  particles hop to the right and to the left at rate $p$ on empty sites, respectively. In addition, particles  change direction at a rate $\alpha$. Consider the sequence of configuration ${\cal C}_1$ to ${\cal C}_4$:
\begin{center}
\begin{tikzpicture}[scale=.9]
\def\H{1.25cm}
\def\L{7cm}
\draw[thick](0,0) node[anchor=east] {${\cal C}_1$} -- (4,0);
\foreach \x in {0,1,2,3}
{\draw (\x+.5, -.15) -- (\x+.5,.15);}
\filldraw[red!40!white](3.5,.5) circle (.2) node {\color{black} $\boldsymbol{-}$};
\filldraw[blue!40!white](0.5,.5) circle (.2) node {\color{black} $\boldsymbol{+}$};

\begin{scope}[yshift=-\H]
\draw[thick](0,0) node[anchor=east] {${\cal C}_2$} -- (4,0);
\foreach \x in {0,1,2,3}
{\draw (\x+.5, -.15) -- (\x+.5,.15);}
\filldraw[red!40!white](2.5,.5) circle (.2) node {\color{black} $\boldsymbol{-}$};
\filldraw[blue!40!white](0.5,.5) circle (.2) node {\color{black} $\boldsymbol{+}$};
\end{scope}

\begin{scope}[yshift=-2*\H]
\draw[thick](0,0) node[anchor=east] {${\cal C}_3$} -- (4,0);
\foreach \x in {0,1,2,3}
{\draw (\x+.5, -.15) -- (\x+.5,.15);}
\filldraw[red!40!white](2.5,.5) circle (.2) node {\color{black} $\boldsymbol{-}$};
\filldraw[blue!40!white](1.5,.5) circle (.2) node {\color{black} $\boldsymbol{+}$};
\end{scope}

\begin{scope}[yshift=-3*\H]
\draw[thick](0,0) node[anchor=east] {${\cal C}_4$} -- (4,0);
\foreach \x in {0,1,2,3}
{\draw (\x+.5, -.15) -- (\x+.5,.15);}
\filldraw[red!40!white](2.5,.5) circle (.2) node {\color{black} $\boldsymbol{-}$};
\filldraw[red!40!white](1.5,.5) circle (.2) node {\color{black} $\boldsymbol{-}$};
\end{scope}

\if{
\begin{scope}[yshift=-4*\H]
\draw[thick](0,0) node[anchor=east] {${\cal C}_5$} -- (4,0);
\foreach \x in {0,1,2,3}
{\draw (\x+.5, -.15) -- (\x+.5,.15);}
\filldraw[red!40!white](2.5,.5) circle (.2) node {\color{black} $\boldsymbol{-}$};
\filldraw[red!40!white](0.5,.5) circle (.2) node {\color{black} $\boldsymbol{-}$};
\end{scope}}\fi


\begin{scope}[xshift=\L]
\draw[thick](0,0) node[anchor=east] {$\overline{\cal C}_1$} -- (4,0);
\foreach \x in {0,1,2,3}
{\draw (\x+.5, -.15) -- (\x+.5,.15);}
\filldraw[blue!40!white](3.5,.5) circle (.2) node {\color{black} $\boldsymbol{+}$};
\filldraw[red!40!white](0.5,.5) circle (.2) node {\color{black} $\boldsymbol{-}$};
\end{scope}

\begin{scope}[yshift=-\H,xshift=\L]
\draw[thick](0,0) node[anchor=east] {$\overline{\cal C}_2$} -- (4,0);
\foreach \x in {0,1,2,3}
{\draw (\x+.5, -.15) -- (\x+.5,.15);}
\filldraw[blue!40!white](2.5,.5) circle (.2) node {\color{black} $\boldsymbol{+}$};
\filldraw[red!40!white](0.5,.5) circle (.2) node {\color{black} $\boldsymbol{-}$};
\end{scope}

\begin{scope}[yshift=-2*\H,xshift=\L]
\draw[thick](0,0) node[anchor=east] {$\overline{\cal C}_3$} -- (4,0);
\foreach \x in {0,1,2,3}
{\draw (\x+.5, -.15) -- (\x+.5,.15);}
\filldraw[blue!40!white](2.5,.5) circle (.2) node {\color{black} $\boldsymbol{+}$};
\filldraw[red!40!white](1.5,.5) circle (.2) node {\color{black} $\boldsymbol{-}$};
\end{scope}

\begin{scope}[yshift=-3*\H,xshift=\L]
\draw[thick](0,0) node[anchor=east] {$\overline{\cal C}_4$} -- (4,0);
\foreach \x in {0,1,2,3}
{\draw (\x+.5, -.15) -- (\x+.5,.15);}
\filldraw[blue!40!white](2.5,.5) circle (.2) node {\color{black} $\boldsymbol{+}$};
\filldraw[blue!40!white](1.5,.5) circle (.2) node {\color{black} $\boldsymbol{+}$};
\end{scope}
\if{
\begin{scope}[yshift=-4*\H,xshift=\L]
\draw[thick](0,0) node[anchor=east] {$\overline{{\cal C}}_5$} -- (4,0);
\foreach \x in {0,1,2,3}
{\draw (\x+.5, -.15) -- (\x+.5,.15);}
\filldraw[blue!40!white](2.5,.5) circle (.2) node {\color{black} $\boldsymbol{+}$};
\filldraw[blue!40!white](0.5,.5) circle (.2) node {\color{black} $\boldsymbol{+}$};
\end{scope}}\fi
\end{tikzpicture}
\end{center}
Starting from ${\cal C}_1$ at $t=0$, let us call $t_i$ the transition times from ${\cal C}_i$ to ${\cal C}_{i+1}$. 
Denoting $i_{1,2},\sigma_{1,2}$ the positions and signs of both particles, the probability density to observe this trajectory during $[0,t_f]$ is given by
\begin{equation}\label{eq:probaforward}
    P[i_{1,2}(t),\sigma_{1,2}(t)]= e^{-2(p+\alpha) t_1} p \times e^{-2( p+\alpha) (t_2-t_1)} p \times  e^{-2\alpha (t_3-t_2)}\alpha \times e^{-(p+2\alpha) (t_f-t_3)} \;.
\end{equation}
The first three factors correspond to the transition ${\cal C}_{i}$ to ${\cal C}_{i+1}$ with $i=1,2,3$ whereas the last factor is the probability of staying in ${\cal C}_4$ up to $t_f$. 
The probability to observe the reverse trajectories $P[i_{1,2}(t_f-t),\sigma_{1,2}(t_f-t)]$ trivially vanishes: a \pluspart{} particles cannot hop backwards in this model, nor can a \minuspart{} hop forward. More interestingly, one may wonder whether reversibility survives if only the positions of the particles are recorded. The converse sequence of positions could indeed be realized by flipping the orientations of the particles (using a so-called kinematic reversibility~\cite{mallmin2019exact}). This corresponds to the sequence $\overline{\cal C}_4$ to $\overline{\cal C}_1$ described above. The probability density of this trajectory can also be evaluated, yielding
\begin{equation}\label{eq:probabackward}
    P[i_{1,2}(t_f-t),-\sigma_{1,2}(t_f-t)]=
         e^{-(p+2\alpha) (t_f-t_3)}\alpha \times 
     e^{-2(p+\alpha) (t_3-t_2)} p \times 
     e^{-2( p+\alpha) (t_2-t_1)}p \times 
    e^{-2( p+\alpha) t_1} 
\end{equation}
Interestingly, the probability densities~\eqref{eq:probaforward}
and~\eqref{eq:probabackward} are unequal, showing the system to lack
TRS in position space at the level of individual trajectories. By
comparing the various terms, the irreversibility is seen to stem from
the different rates at which the system hops out of configurations
${\cal C}_3$ and $\overline{\cal C}_3$ ($2\alpha$ and $2\alpha+2p$,
respectively). These different escape rates lead to the factors
$e^{-2(p+\alpha)(t_3-t_2)}$ and $e^{-2\alpha(t_3-t_2)}$, which are the
sole terms violating TRS. If typical trajectories include an extensive
number of collisions in time, this will lead to a non-zero entropy
production rate. All in all, it is thus the interaction between the
particles which generates the violation of TRS, as first explained
in~\cite{mallmin2019exact}.
\end{tcolorbox}

\subsection{TRS and effective equilibrium descriptions of MIPS.}

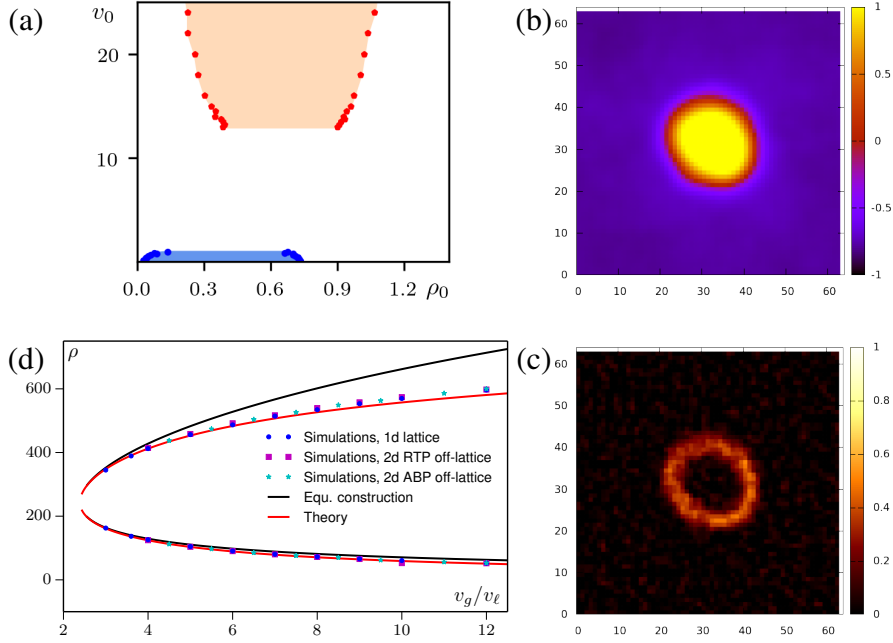
\begin{figure}
    \centering
    \begin{tikzpicture}
      \path(0,0) node {\includegraphics[totalheight=4cm]{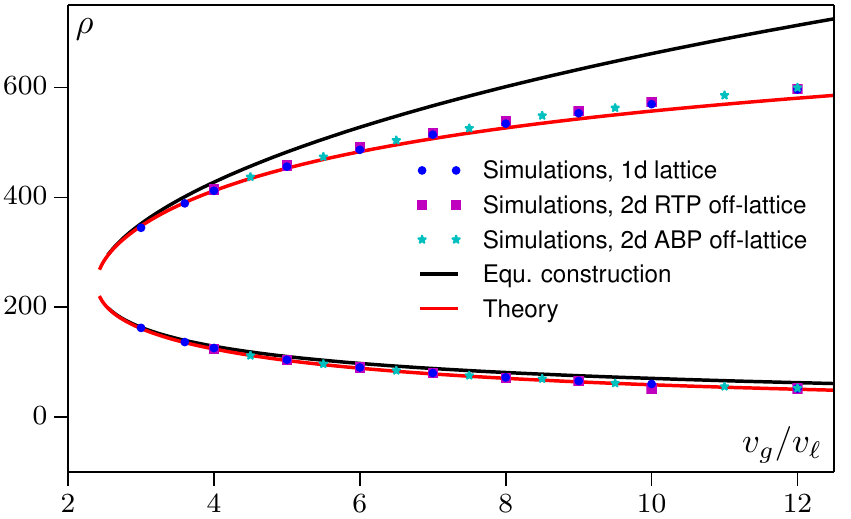}};
      \path(6,4.5) node {\includegraphics[totalheight=4cm]{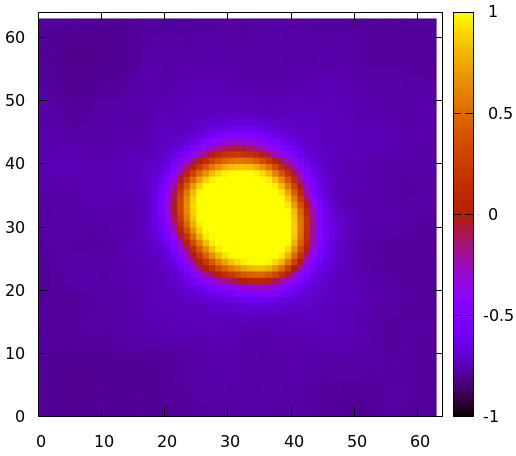}};
      \path(6,0) node {\includegraphics[totalheight=4cm]{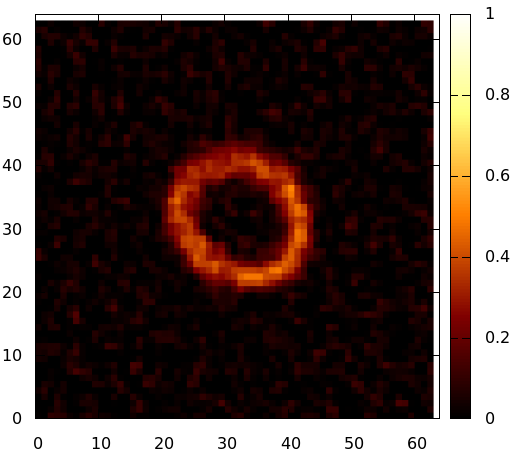}};
      \path(0,4.5) node {\includegraphics[totalheight=4cm]{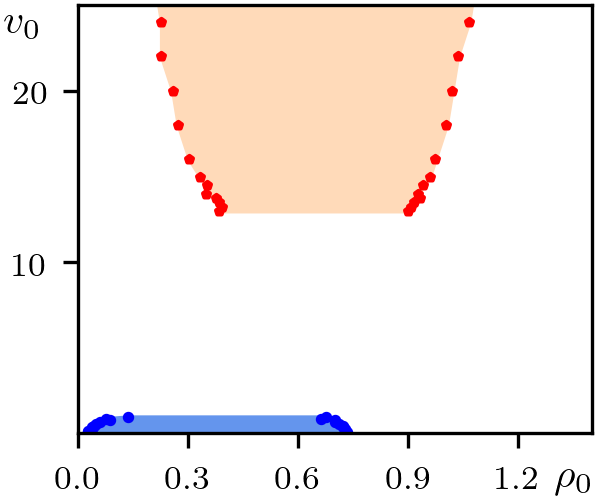}};
      \draw(-3.25,1.7) node {(d)};
      \draw(-3.25,6.2) node {(a)};
      \draw(3.5,1.7) node {(c)};
      \draw(3.5,6.2) node {(b)};
    \end{tikzpicture}
    \caption{{\bf (a)} Phase diagram of ABPs interacting via a
      Lennard-Jones potential. For vanishing self-propulsion $v_0$,
      the system is effectively in equilibrium and undergoes a
      traditional liquid-gas phase separation. The phase-separated
      (blue) region survives at small $v_0$ until the self-propulsion
      is strong enough to overcome attractive interactions, hence
      vaporising the liquid phase. A reentrance into a phase-separated
      (orange) region is observed at larger $v_0$. The latter stems
      from a motility-induced phase separation and would also be
      observed in the absence of the attractive tail of the pair
      potential. Symbols correspond to binodals measured in
      simulations of self-propelled particles interacting via a
      Lennard-Jones potential. The phase-separated regions are
      estimated numerically up to unresolved critical regions using a
      methodology described in Appendix~\ref{app:LJ}. {\bf (b) \& (c)}
      Simulations of a scalar field-theory undergoing motility-induced
      phase separation. The order parameter $\Phi(x,y)$ shown in panel
      (b) distinguishes a liquid phase ($\Phi=1$) from a gas one
      ($\Phi=-1$). The entropy production rate can be decomposed
      spatially as $\sigma=\int d^2 {\bf r} \hat \sigma({\bf r})$. The
      plot of $\hat \sigma({\bf r})$ in panel (c) shows the interface
      to contribute most to the TRS violations measured by
      $\sigma$. Adapted
      from~\protect\cite{Nardini:PhysRevX.7.021007}. {\bf (d)} Phase
      diagrams of microscopic models of quorum-sensing active
      particles (symbols). Equating the generalized pressures $P$ and
      chemical potentials $\mu$ in coexisting phases lead to a
      theoretical prediction (red line) in quantitative agreement with
      microscopic simulations. An equilibrium construction based on a
      local approximation of $\mu$ (black solid line) is
      quantitatively incorrect. Adapted
      from~\protect\cite{solon2018generalized}. }
    \label{fig:MIPS}
\end{figure}

Consider an equilibrium system comprising Brownian particles
interacting via a pairwise potential made of a hard core and a soft
attractive tail. The steady state of the system is found by balancing
entropy, which favors disordered configurations, with energy. At
moderate densities, the latter favors cohesion thanks to the
attractive part of the potential. As the temperature is lowered,
energy wins over entropy and a gas-to-liquid phase transition is
observed. When the total number of particles is conserved, this leads
to phase coexistence. Now, make the particle active. A natural way to
preserve phase separation is to make self-propulsion weak enough that
it does not overcome the attractive interactions. Interestingly,
however, the largest part of the phase diagram exhibiting liquid-gas
coexistence is not found at small self-propulsion speeds but at large
ones~\cite{redner2013reentrant}, and it is observed even in the
absence of any attractive
tail~\cite{Fily:2012:PRL,redner2013structure} (See
Fig.~\ref{fig:MIPS}). A cohesive phase may thus emerge out of purely
repulsive pairwise forces in active systems.  This phase-separated
state cannot be accounted for using the Boltzmann weight and is a
trademark of active particles. It results from a motility-induced
phase separation~\cite{Cates:2015:ARCMP} through which particles whose
self propulsion is hindered at high density separate between a dense,
almost arrested, phase and a dilute active gas.

Despite its non-Boltzmann nature, the dynamics of such a phase-separated system at the macroscopic scale, however, does not reveal any clear breakdown of TRS  at first sight, and the difference with an equilibrium phase separation is hard to pinpoint. Therefore,
many ideas based on equilibrium theories have been proposed to account for MIPS~\cite{Tailleur:2008:PRL,Speck2014PRL,takatori2015towards,redner2016classical,paliwal2017non,paliwal2018chemical,Solon2018PRE,solon2018generalized}. Below we first focus on large-scale TRS violations in MIPS before discussing when and how a generalization of equilibrium approaches may be used to describe MIPS and other phase transitions.

\paragraph{Large-scale TRS violations in MIPS.}
We now focus on the MIPS observed in systems with quorum-sensing interactions, i.e. when the speed of a particle explicitly depends on the local density of particles around it~\cite{Tailleur:2008:PRL,lavergne2019group,bauerle2018self}:
\begin{equation}\label{eq:QS}
    \dot {\bfr_i} = v(\bfr_i,[\rho]) \bfu_i\;,\qquad \rho(\bfr) \equiv \sum_i \delta(\bfr-\bfr_i)\;,
\end{equation}
where $\bfu_i$ undergoes either rotational diffusion or instantaneous Poisson-distributed tumbles. A common form for $v(\bfr_i,[\rho])$ is then~\cite{Solon2015EPJST,solon2018generalized}
\begin{equation}\label{eq:kernel}
    v(\bfr_i,[\rho]) = f[ \tilde \rho(\bfr,t)] \;,\qquad \tilde\rho(\bfr,t)\equiv  \int d\bfr \rho(\bfr) K(\bfr_i-\bfr)\;,
\end{equation}
where $K$ is a kernel which can stem from integrating out the signal
used by the particles to communicate~\cite{o2020lamellar}, and $f$
determines how the particles react to their
peers. Equations~\eqref{eq:QS} and~\eqref{eq:kernel}, together with
$K(\bfr)=\delta(\bfr)$ and $f$ a linearly decaying function, have been
proposed as a toy model for systems with pairwise repulsive
forces~\cite{Thompson:2011:JSM,bialke2013microscopic}, even though the
latter offer a richer phenomenology such as hexatic
order~\cite{klamser2018thermodynamic,digregorio2018full,de2019active}
or bubbly phases~\cite{tjhung2018cluster,shi2020self}.

A glimpse into the large-scale behaviour of this system can be
obtained using a diffusive approximation on Eq.~\eqref{eq:QS}, which
leads to the fluctuating
hydrodynamics~\cite{Solon2015EPJST,solon2018generalized,mahdisoltani2019controlled}
\begin{equation}\label{eq:flucthyd}
    \dot \rho  = \nabla \cdot [ \rho D \nabla \mu + \sqrt{2 \rho D} \boldsymbol\Lambda]\quad\text{with}\quad D=\frac{v^2 \tau}{d}\quad\text{and}\quad \mu=\log \rho+\log f(\tilde \rho)\;,
\end{equation}
where $\boldsymbol\Lambda$ is a Gaussian white-noise field of zero
mean and correlations
$\langle\Lambda_\alpha(\bfr,t)\Lambda_\beta(\bfr',t')\rangle=\delta_{\alpha\beta}\delta(\bfr-\bfr')\delta(t-t')$
and $d$ is the number of space dimensions. By analogy to the
equilibrium model $B$, we refer to $\mu$ as a chemical potential. The
dynamics~\eqref{eq:flucthyd} satisfies TRS if and only if $\mu(\bfr)$
is a gradient of a free-energy functional~\cite{Solon2015EPJST}. This
holds when a generalization of the Schwarz equality to functional
derivatives is obeyed:
\begin{equation}\label{eq:integrability}
    \frac{\delta \mu(\bfr,[\rho])}{\delta \rho(\bfr')}=\frac{\delta \mu(\bfr',[\rho])}{\delta \rho(\bfr)}\;,
\end{equation}
where the equality is understood as between distributions~\cite{o2020lamellar}. 

For a generic $f(\tilde \rho)$, Eq.~\eqref{eq:integrability} is not
satisfied and the dynamics~\eqref{eq:flucthyd} does not obey TRS. An
interesting exception is when $K(\bfr)=\delta(\bfr)$; $v$ is then a
local function of the density and Eq.~\eqref{eq:flucthyd} satisfies an
exact TRS~\cite{Tailleur:2008:PRL}. For more general kernels, a
free-energy functional is only found for $f(\tilde \rho)=v_0
\exp(-\lambda \tilde \rho)$~\cite{grafke2017spatiotemporal}.  When
this is not the case, it is interesting to see which features of the
dynamics are directly responsible for breaking TRS. To do this, we can
expand $\mu$ to second order in gradients:
\begin{equation}\label{eq:mu2nd}    \mu[\bfr]=\log\rho+\log f(\rho)-\kappa\Delta \rho \quad\text{where}\quad \kappa(\rho)=-\ell^2 \frac{f'(\rho)}{f(\rho)}\;,
\end{equation}
and $\ell^2=\int d \bfr  K(\bfr)\bfr^2$ is a measure of the interaction range~\cite{solon2018generalized}. The entropy production rate can then be computed using methods akin to those presented in~\cite{Nardini:PhysRevX.7.021007} as:
\begin{equation}\label{eq:sigmaQSAPs}
\sigma = - \lim_{t\to\infty} \frac 1 {2t} \int_0^t ds \int d\bfr \dot\rho(\bfr,t) \kappa[\rho(\bfr,t)]  \Delta \rho +{\cal O}(\nabla^4) \simeq \frac 1 {2} \int d\bfr \langle \dot\rho(\bfr) \kappa[\rho(\bfr)]  \Delta \rho\rangle  \;,
\end{equation}
where the last equality stems from ergodicity.  A few comments are in
order. First, a completely local approximation to $\mu$, that neglects
all gradient terms, leads to a vanishing entropy production rate. This
is consistent with the apparent macroscopic similarity between MIPS
and an equilibrium liquid-gas separation. Then, taking gradients into
account, TRS is broken and Eq.~\eqref{eq:sigmaQSAPs} offers a spatial
decomposition of the entropy production rate. It suggests that the
latter is most pronounced in inhomogeneous regions, namely next to
interfaces between coexisting phases. Figure~\ref{fig:MIPS} shows a
measure of $\sigma$ in a simplified version of the
field-theory~\eqref{eq:flucthyd} undergoing MIPS, which is commonly
referred to as active model B. The entropy production rate is indeed
peaked at the interface between the gas and liquid phases, a result
that was confirmed using microscopic simulations of
AOUPs~\cite{martin2020statistical}. This role of interfaces in TRS
breakdown also has consequences for static properties, as exemplified
by the construction of the phase diagram which we now discuss.

\paragraph{Phase diagram and generalized thermodynamics of MIPS.}
A natural question is whether the dynamical differences between MIPS
and an equilibrium phase separation also affect the static properties,
and in particular the density profiles connecting coexisting phases
which, as we discuss below, determine the phase diagram.  The simplest
way of answering this question is to take a mean-field
approximation. In a homogeneous system, we can use a local
approximation of $\mu$, defined in Eq.~\eqref{eq:flucthyd}, to obtain the
Landau free energy ${\cal F}=\int d\bfr \phi[\rho(\bfr)]$, where
$\phi(\rho)= \rho (\log \rho-1)+F(\rho)$ with $F'(\rho)=\ln
f(\rho)$. The linear stability of a homogeneous phase at density
$\rho_0$ is then lost whenever $\phi''(\rho_0)<0$, i.e.
$\rho_0 f'(\rho_0)<-f(\rho_0)$. This defines a spinodal region within
which the system is linearly unstable to
MIPS~\cite{Tailleur:2008:PRL}. At this level, there is no difference
with an equilibrium mean-field theory. To account for the resulting
phase-separated state, it is necessary to consider inhomogeneous
profiles.

To do so, we again use the leading-order gradient expansion of $\mu$
given by Eq.~\eqref{eq:mu2nd}. At this level, the mapping to equilibrium
is violated and equation~\eqref{eq:integrability} is not
satisfied. (Again, an exception occurs for
$f(\tilde \rho)=v_0 \exp(-\lambda
\tilde\rho)$~\cite{Tailleur:2008:PRL}.)  The
attempts~\cite{Tailleur:2008:PRL,Speck2014PRL,takatori2015towards} to
build the phase diagram using common-tangent constructions on
$\Phi(\bfr)$ are thus bound to
fail~\cite{wittkowski2014scalar,Solon:2015:PRL}. Somewhat
surprisingly, if $\mu(\bfr)$ cannot be written as the functional
derivative of a free energy with respect to $\rho(\bfr)$, a gradient
form can be found upon a change of variable $\rho \to R(\rho)$, where
$ R'(\rho) = 1/\kappa(\rho)$. Then,
$\mu(\bfr)=\nabla \frac{\delta {\cal H}}{\delta R(\bfr)}$, where
${\cal H}=\int d \bfr [\Phi(R)+\frac{\kappa}{2 R'}(\nabla R)^2]$ and
$\Phi'(R)=\log \rho(R)+\log f(\rho(R))$. This allows identifying $\mu$
and a generalized pressure $P\equiv R \frac{d \Phi}{d R}-\Phi$ as
state variables, which are equal in coexisting
phases~\cite{Solon2018PRE,solon2018generalized}.  Equivalently, a
common-tangent construction on $\Phi(R)$ can be used to construct the
coexisting densities. Despite its mean-field nature, this construction
was shown to give precise estimates of the phase diagram for 
microscopic models of quorum-sensing active particles undergoing MIPS,
without any fitting
parameters~\cite{Solon2018PRE,solon2018generalized} (See
Fig~\ref{fig:MIPS}).  Note that the agreement does not extend up to
the putative critical point close to which the mean-field theory is
expected to break-down in a Ginzburg interval (which is not
numerically resolved). Furthermore, at large propulsion speeds, domain
walls become so sharp that higher-order gradients cannot be neglected
leading to quantitative differences with the predictions
of~\eqref{eq:mu2nd}.

\subsection{Tactic dynamics.} In the case of MIPS, the strong TRS violations exhibited at the single particle level ($\sigma$ is infinite in the absence of translational diffusion) are thus partially erased upon coarse-graining. TRS violations, occuring mostly at interfaces, survive because condition~\eqref{eq:integrability} is generically not satisfied. A natural question is then whether there exist active systems whose fluctuating hydrodynamics satisfy the integrability condition~\eqref{eq:integrability}. This question was answered for tactic dynamics, in which particles move according to Eq.~\eqref{eq:dynRTP} and whose orientations $\bfu_i$ perform tumbles at rate $\alpha$ and diffuse with rotational diffusivity $\Gamma$. Taxis is implemented by coupling self propulsion with the gradient of a  field $c(\bfr)$ through~\cite{saha2014clusters,o2020lamellar,mahdisoltani2019controlled}
\begin{equation}\label{eq:chemospeed}
v_p = v_0 - v_1 \mathbf{u}_i\cdot \nabla c\qquad \alpha = \alpha_0 + \alpha_1 \mathbf{u}_i\cdot \nabla c \qquad \Gamma = \Gamma_0 + \Gamma_1\mathbf{u}_i\cdot \nabla c
\end{equation}
The field $c(\bfr)$ can be either externally imposed or, more interestingly, can be a functional of the particle density. Under the hypothesis that the large-scale diffusive scaling of non-interacting active particles survives the addition of interactions, the fluctuating hydrodynamics of this model can be constructed and Eq.~\eqref{eq:integrability} can be satisfied in a number of non-trivial cases~\cite{o2020lamellar}. In particular, when $c(\bfr)$ is a diffusive signalling field produced by the particles, this system models swimming bacteria interacting via chemotaxis. The integrability of the large-scale dynamics then reveals a mapping between bacteria interacting both via chemotractant and chemorepellent and Brownian colloids interacting via attractive and repulsive forces. While the model violates TRS at the microscopic scale, the latter is restored upon coarse-graining and the rich phase diagram of this system can be accounted for exactly using equilibrium theories. Note that the fate of TRS at the fluctuating hydrodynamic level in active systems is an open, non-trivial question. For instance, another model of tactic dynamics defined as $\dot \bfr_i= \nu_1 \nabla c + \nu_2 (\bfu(\theta) \cdot \nabla) \nabla c$ was shown to lead to a violation of TRS at the coarse-grained level~\cite{mahdisoltani2019controlled}. The results of~\cite{o2020lamellar} thus establish a non-trivial embedding of equilibrium physics in active matter systems, but slight perturbations, albeit leading to minor modifiations of large-scale behaviours, may trigger TRS violations.

\subsection{Small $\tau$ regime.} Finally, let us mention that the equilibrium regime recovered in the $\tau\to0$ limit while keeping $kT_{\rm eff}=v_0^2 \tau/\mu$ constant discussed for the single-particle case survives under the addition of pairwise forces. The departure from this limiting case has been studied analytically for AOUPs, both using Markovian approximations~\cite{wittmann2017effective1,wittmann2017effective2} and an explicit small-$\tau$ expansion~\cite{Fodor:2016:PRL,martin2020statistical}. To linear order in $\tau$, the system satisfies detailed balance with a non-Boltzmann form. Casting $\ln P$ into an effective potential, pairwise repulsive forces lead to effective attractive interactions~\cite{farage2015effective}. While the latter do not allow to quantitatively account for MIPS~\cite{rein2016applicability}, they qualitatively capture how self-propulsion turns repulsive forces into attractive ones. Finally, note that \textit{bona fide} equilibrium systems can be studied under the perturbation of a small self-propulsion velocity, which allows studying the fate of thermodynamical concepts like surface tension or chemical potential in weakly active systems~\cite{paliwal2017non,paliwal2018chemical}. The same applies to equilibrium phases, like the hexatic phase observed in 2D melting, which has been shown to survive activity~\cite{klamser2018thermodynamic,klamser2019kinetic,digregorio2018full}.

\section{Perspectives}
Most active-matter models such as Eq.~\eqref{eq:dynact} aim at
modelling the effective dynamics of self-propelled particles and not
the irreversible consumption of energy powering the propulsion
force. As such, they are unable to account for all the dissipative
processes occuring in the system, hence rendering the theoretical
assessment of its full irreversibility a somewhat hopeless endeavour.
This highlights an important property of nonequilibrium systems:
irreversibility depends on the degrees of freedom under study and on
the scale of their description. This raises the question of the scale
at which dissipation is maximal. It also opens up the possibility
that equilibrium statistics prove relevant to describe specific scales
or observables. Effective-equilibrium descriptions have indeed proven
successful in qualitatively accounting for part of active matter
phenomenology such as the motility-induced phase separation. On the
contrary, time is now ripe for the identification of TRS-violating
emerging behaviours and for the quantification of entropy-production rates
in active pattern-forming systems. In turn, the connection between
entropy production rates and more directly accessible observables such
as steady-state currents remains an open challenge at the
coarse-grained scale. These questions are particularly relevant in the
biophysical context that gave birth to the field of active matter:
cells are constantly dissipating energy to exert biological functions;
assessing which of them genuinely rely on nonequilibrium processes is
a fascinating open challenge.

Beyond the violation of TRS observed in active systems, we have
reviewed their most salient non-Boltzmann features. That the
steady-state distributions  are not captured by the
Boltzmann weight indeed allows for phenomenologies unmatched in passive
systems, but it also impairs our ability to design smart active
materials. An important challenge facing the community is then to
develop alternatives to equilibrium thermodynamics that would grant us
the same level of intuition and control over active systems as we have
over passive ones. Natural starting points are then the various limits
in which effective equilibrium concepts can be proven relevant. From
the small-but-non-zero persistence-time limit, to emerging TRS
restored by coarse-graining, to the identification of state functions,
we have reviewed cases in which efforts were rewarded. Beyond these
cases, this calls for an extensive exploration of the fate of
thermodynamic state variables (entropy, chemical potentials,
\textit{etc}.). At the local level, inferring effective interactions
by means of rapidly developing learning algorithms seems a promising
avenue~\cite{turci2021phase,bag2020interaction,cichos2020machine,colen2020machine}.

\textit{Acknowledgment:} JO, JT, FvW acknowledge support from ANR
grant THEMA; YK acknowledges support from the ISF and from an NSF-BSF
grant. We thank Massimo Pica Ciamarra and Yanwei Li for sharing the
data published in~\cite{li2020phase} that were used to calibrate the
simulations shown in Fig.~\ref{fig:MIPS}a. The authors benefited from
participation in the 2020 KITP program on Active Matter supported by
the Grant NSF PHY-1748958.

\appendix

\section{Entropy production}
\label{app:ep}
The forward trajectory $\bfr(t)$ and $\theta(t)$ uniquely characterize
the realization of the noise through
\begin{equation}
  \bfeta(t)=\frac{  \dot \bfr(t) - v_0 \bfu(\theta(t))}{\sqrt{2 D}}
\end{equation}
For the reversed trajectory $\bfr^{\rm r}(t)=\bfr(t_f-t)$,
$\theta^{\rm r}(t)=\theta(t_f-t)$ to be observed, the surrounding fluid
molecules have to produce a different noise
\begin{equation}\label{eq:reversednoise}
  \bfeta^{\rm r}(t)=\frac{  \dot \bfr^{\rm r}(t) - v_0 \bfu(\theta^{\rm r}(t))}{\sqrt{2 D}}=\frac{  -\dot \bfr(t_f-t) - v_0 \bfu(\theta(t_f-t))}{\sqrt{2 D}}=-\bfeta(t_f-t)-\frac{2 v_0 \bfu(\theta(t_f-t))}{\sqrt{2 D}}
\end{equation}
To characterize the irreversibility of the process it is useful to
note that the probability of a trajectory is given by its
Onsager-Machlup form~\cite{onsager1953fluctuations,woillez:2019:PRL}
\begin{equation}
  P[\{\bfr(t),\theta(t)\}]=Z^{-1} P[\{\theta(t)\}|\theta_0] \exp\Big[-\frac 1 {4D} \int_0^{t_f}dt [\dot \bfr(t)-v_0 \bfu(\theta)]^2\Big] P_0(\bfr_0,\theta_0)
\end{equation}
where $P[\{\theta(t)\}|\theta_0]$ is the probability of the
realization of $\theta(t)$ starting from $\theta_0$, $P_0$ is the
probability of the initial condition, and $Z^{-1}$ is a
normalization. Note that, for fully randomizing tumbles,
$P[\{\theta(t_f-t)\}|\theta_f]=P[\{\theta(t)\}|\theta_0]$. Taking the
logarithm of the ratio between forward and backward trajectories yield
the corresponding `path-wise entropy
production'~\cite{seifert2005entropy}:
\begin{equation}
\hat \Sigma [\{\bfr(t),\theta(t)\}] \equiv  \log \frac{P[\{\bfr(t),\theta(t)\}]}{P[\{\bfr(t_f-t),\theta(t_f-t)\}]} = \frac{\mu}{D} \int_0^{t_f} dt  \dot \bfr \cdot \bfp +\log\frac{P_0(\bfr_0,\theta_0)}{P_f(\bfr_f,\theta_f)},
\end{equation}
where $P_f(\bfr_f,\theta_f)$ is the probability of being at
$\bfr_f\equiv\bfr(t_f)$ and $\theta_f\equiv \theta(t_f)$ given that
the initial condition was sampled according to $P_0$.

\section{Self-consistency of the effective temperature regime}
\label{app:selfcontpot}

Let us consider the self-consistency of Eq.~\eqref{eq:Boltzmanneff} in
the presence of a confining potential. Close to the minimum $x_0$ of
the potential, the condition $v_0 \gg \mu V'(x)$ trivially
holds. Let us expand the distribution~\eqref{eq:Boltzmanneff} around
$x_0$. It is then given by a locally  Gaussian distribution
\begin{equation}
  P(x) \sim \exp\left[-\frac{V''(x_0)(x-x_0)^2}{2 T_{\rm eff}}\right]
\end{equation}
This predicts a typical displacement
$x_t \sim \sqrt{T_{\rm eff}/V''(x_0)}$ which in turn leads to a
typical force that scales as
$V'(x_t)\simeq \sqrt\frac{\tau v_0^2 V''(x_0)}{\mu}$, where we have
used $T_{\rm eff} = v_0^2\tau/\mu$. The condition $v_0 \gg \mu V'(x)$
then becomes $\mu V''(x_0) \tau \ll 1$: the typical time between two
tumbles has to be much shorter than the relaxation time inside the
potential well.

\section{Non-locality of the steady-state distribution}
\label{app:nonlocal}
Consider the effective potential given in Eq.~\eqref{eq:AOUPST}. Its
non-local nature becomes apparent if one adds a localized perturbation
$\delta V(x)=\epsilon \delta(x-y)$, centered around $y$, to the
potential $V(x)$. The effective potential $V_{\rm eff}(x)$ then picks
up a contribution due to $\delta V$, because of the term
$\frac{\int^x V'(y)^2 V^{(3)}(y)dy}2$. To linear order in $\epsilon$,
it reads
\begin{equation}
  \delta V_{\rm eff}(x)= \frac{\epsilon \tau^2}{2}
[(V'(y)^2)^{(3)}+2V''(y)V^{(3)}(y)+2V'(y)V^{(4)}(y)  ]\Theta(x-y)\;,
\end{equation}
which adds a global step to the density profile at $x=y$.

\section{Self-propelled ABPs interacting via a Lennard-Jones potential}
\label{app:LJ}
We consider $N$ active Brownian particles evolving in two space dimensions under the dynamics:
\begin{equation}
  \dot{\bf r}_i=v_0{\bf u}(\theta_i) -\sum_j \nabla V({\bf r}_i-{\bf r}_j)+\sqrt{2 D_t} \boldsymbol{\eta}_i\qquad \dot \theta_i = \sqrt{2 D_r} \xi_i
\end{equation}
where $\boldsymbol{\eta}_i$ and $\xi_i$ are independant Gaussian white
noises, and $V$ is the Lennard-Jones potential
\begin{equation}
  V({\bf r})=4 \varepsilon \left(\frac{\sigma^{12}}{{\bf r}^{12}}-\frac{\sigma^{6}}{{\bf r}^{6}}\right)\;.
\end{equation}
Interactions between particles are truncated at $|{\bf r}|=2.7\sigma$
and we used periodic boundary conditions. Figure~\ref{fig:MIPS}a was
obtained using $\sigma=0.37$, $T=0.4$ and $D_r=2$. The phase-separated
regions were found by running simulations for $\rho_0=2.5\sigma^2$
($v_0<8$) and $\rho_0=4\sigma^2$ ($v_0>8$) in systems of line sizes
$L=80$ up to $t=2\,000$, where $\rho_0$ is the rescaled number density
$\rho_0\equiv N \sigma^2/L^2$. The symbols correspond to estimates of
the coexisting densities. The latter were obtained from the local
maxima of histograms of the density field, constructed using bins of
linear size $4$.

\vspace{.5cm}
\bibliographystyle{plain}
\bibliography{biblio.bib}

\end{document}